\documentstyle[epsfig,axodraw,12pt]{article}

\newcommand{\slashs}[1]{\not{\!#1}}
\begin{document}
\vskip 0.5cm
\renewcommand{\thefootnote}{\fnsymbol{footnote}}
\newcommand{\cita} [1] {$^{\hbox{\scriptsize \cite{#1}}}$}
\newcommand{\prepr}[1] {\begin{flushright}  {\bf #1} \end{flushright}}
\newcommand{\titul}[1] {\begin{center}{\Large {\bf #1 } }\end{center}}

\newcommand{\autor}[1] {\begin{center}{\bf \lineskip .3cm #1} \end{center}}

\newcommand{\address}[1] {\begin{center}  {\normalsize \bf \it #1 }
\end{center}}
\newcommand{\abstr}[1] {{\begin{center} \vskip .5cm {\bf \large Abstract
                        \vspace{0pt}} \end{center}}\begin{quote} \small #1
                        \end{quote}}

\begin{titlepage}
\titul{\bf Nonfactorizable contributions to $B$ meson decays into
charmonia}

\autor{Chuan-Hung Chen$^1$\footnote{chchen@mail.ncku.edu.tw} and
Hsiang-nan Li$^{1,2}$\footnote{hnli@phys.sinica.edu.tw}}

\vskip2.0cm

\address{$^1$Department of Physics, National
Cheng-Kung University, \\
Tainan, Taiwan 701, Republic of China}

\address{$^2$Institute of
Physics, Academia Sinica, Taipei, Taiwan 115, Republic of China}

\vskip 1.0cm
\begin{abstract}
We analyze the $B\to (J/\psi, \chi_{c0},\chi_{c1},\eta_c)K^{(*)}$
decays, which have small or even vanishing branching ratios in
naive factorization assumption (FA). We calculate nonfactorizable
corrections to FA in the perturbative QCD approach based on $k_T$
factorization theorem. The charmonium distribution amplitudes are
inferred from the non-relativistic heavy quarkonium wave
functions. It is found that the nonfactorizable contributions
enhance the branching ratios and generate the relative phases
among helicity amplitudes of the above modes. Most of the observed
branching ratios, polarization fractions, and relative phases,
except those of $B\to\eta_c K$, are explained. Our predictions for
the $B\to (\chi_{c0},\chi_{c1},\eta_c)K^*$ decays can be compared
with future data.

\end{abstract}

\vskip2.0cm
\end{titlepage}

\section{INTRODUCTION}

It has been known that the naive factorization assumption (FA)
\cite{BSW,fac} does not apply to exclusive $B$ meson decays into
charmonia, such as $B \to J/\psi K$ \cite{a1a2}. These modes
belong to the color-suppressed category \cite{NPe}, for which
predictions from FA are always small due to the vanishing Wilson
coefficient $a_2\sim 0$. However, the branching ratios measured by
Babar recently \cite{Babar-hep04}
\begin{eqnarray}
B(B^+ \to J/\psi K^+) &=& (10.61\pm 0.15 \pm 0.48) \times 10^{-4}
\, ,\nonumber \\
B(B^0 \to J/\psi K^0) &=& (8.69 \pm 0.22 \pm 0.30) \times 10^{-4}
\, , \label{eq:BRexp}
\end{eqnarray}
imply a larger parameter $a_2(J/\psi K)\approx 0.20-0.30$
\cite{a1a2}. The same difficulty has appeared in other similar
decays $B\to (\chi_{c0},\chi_{c1},\eta_c)K$: the observed
branching ratios are usually many times larger than the
expectations from FA. For example, the $B\to\chi_{c0}K$ decays do
not receive factorizable contributions, but the data of their
branching ratios are comparable to those of $B\to J/\psi K$ in
Eq.~(\ref{eq:BRexp}).

\begin{figure}[t!]
\begin{center}
\epsfig{file=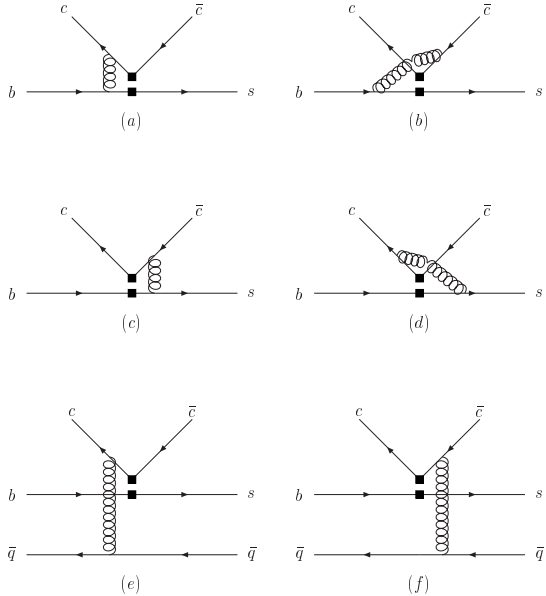, height=3.5in}
\end{center}
\caption{Feynman diagrams for nonfactorizable corrections to
$\overline{B} \rightarrow J/\psi K$.} \label{fvs}
\end{figure}

Many attempts to resolve this puzzle have been made in more
sophisticated approaches (for a review, see \cite{LRev}). The
large $a_2(J/\psi K)$ leads to a natural conjecture that
nonfactorizable contributions, such as the vertex and spectator
corrections from Figs.~\ref{fvs}(a)-\ref{fvs}(d) and from
Figs.~\ref{fvs}(e) and \ref{fvs}(f), respectively, may play an
important role. It has been found \cite{chay} in the QCD-improved
factorization (QCDF) \cite{BBNS} that the above nonfactorizable
contributions, resulting in the branching ratio $B(B^0 \to J/\psi
K^0)\approx 1\times 10^{-4}$,  are too small to explain the data.
Since only the leading-twist (twist-2) kaon distribution amplitude
was included in \cite{chay}, the authors of \cite{Cheng} added the
twist-3 contribution, which is chirally enhanced, though being
power-suppressed in the heavy quark limit. Unfortunately,
Figs.~\ref{fvs}(e) and \ref{fvs}(f) generate logarithmical
divergences from the end-point region, where the parton momentum
fraction of the kaon is small. To make an estimation, arbitrary
cutoffs for parameterizing the divergences, i.e., large
theoretical uncertainties, have been introduced. The end-point
singularities become more serious in the QCDF analysis of the $B
\to (\chi_{c0}, \chi_{c1})K$ decays \cite{Chao2}. It has been
claimed that the data of those modes involving $J/\psi$,
$\chi_{c1}$, and $\eta_c$ are not understandable in QCDF, and only
those involving $\chi_{c0}$ are \cite{chay,Chao2,Chao1}.

The $B\to J/\psi K$ decays have been also studied in light-cone
sum rules (LCSR) \cite{Melic02,WLH03}, and the small branching
ratio from the factorizable contribution
\cite{Melic02},
\begin{eqnarray}
B(B \to J/\psi K)_{\rm fact} \approx 3 \times 10^{-4}\;,
\label{eq:BRnf}
\end{eqnarray}
was confirmed. The nonfactorizable contributions considered in
\cite{Melic02,WLH03} refer, however, to those arising from the
three-parton kaon distribution amplitudes. In the corresponding
diagrams an additional valence gluon from the kaon attaches one of
the charm quarks in the $J/\psi$ meson. Adding these pieces, one
arrived at the parameter $a_2(J/\psi K) \sim 0.14 -0.17$, which is
still insufficient to account for the data. The calculation of
Fig.~\ref{fvs} in LCSR, involving two-loop integrals, has not yet
been performed. Hence, the conclusion in \cite{Melic02,WLH03}
simply indicates that the nonfactorizable contributions to the
$B\to J/\psi K$ decays from higher Fock states of the involved
mesons are negligible.

We doubt that the conclusion from QCDF is solid for two reasons.
First, the end-point singularities render the estimation of the
nonfactorizable contributions out of control. Second, the simple
asymptotic models were adopted for the charmonium distribution
amplitudes, which, without any theoretical and experimental bases,
may underestimate the nonfactorizable contributions. Motivated by
the above observations, we shall investigate the nonfactorizable
corrections to FA from Fig.~\ref{fvs} using the following method.
We take into account the vertex corrections through the variation
of the renormalization scale for the factorizable contributions,
because their evaluation requires more theoretical inputs, which
have not yet been available. The spectator amplitudes will be
computed in the perturbative QCD (PQCD) approach based on $k_T$
factorization theorem \cite{LY1,CL,KLS,LUY}. The formalism,
similar to that of the color-suppressed mode $\bar B^0\to
D^0\pi^0$ \cite{KKLL}, is free from the end-point singularities.
We infer the charmonium distribution amplitudes from the
non-relativistic heavy quarkonium bound-state wave functions,
which have been shown to well explain the cross section of
charmonium production in $e^+e^-$ collisions \cite{BC04}. Based on
the universality of hadron distribution amplitudes, these more
sophisticated models should be able to account for the exclusive
$B$ meson decays into charmonia, if QCD factorization theorem
works.

The $B\to J/\psi K^{(*)}$ decays have been studied in PQCD
\cite{YL,SSU,Chen03}, compared to which the new ingredients of
this work are: 1) we clarify several controversial statements on
the end-point singularities in the $B\to (J/\psi,\chi_{c1}) K$
analysis in the literature; 2) the $K^{(*)}$ meson distribution
amplitudes derived from QCD sum rules \cite{PB1,PB2} are included
up to two-parton and twist-3 level; 3) the models for the
charmonium distribution amplitudes have both theoretical and
experimental bases; 4) not only the $B\to J/\psi K^{(*)}$ decays,
but $B\to(\chi_{c0}, \chi_{c1}, \eta_{c}) K^{(*)}$ are
investigated. It will be shown that the obtained branching ratios,
polarization fractions, and relative phases among various helicity
amplitudes, except those of $B\to\eta_c K$, are all in consistency
with the existing data. The observed $B\to\eta_c K$ branching
ratios, significantly larger than our results, are the only
puzzle. Therefore, our conclusion differs from that drawn in QCDF
\cite{chay,Chao2,Chao1}. The predictions for the $B\to
(\chi_{c0},\chi_{c1},\eta_c)K^*$ modes can be compared with future
measurements.

We present our formalism for the $B\to J/\psi K^{(*)}$ decays as
an example in Sec.~II. The factorizable contributions are treated
in FA, since the $B\to K^{(*)}$ transition form factors in the
present case are characterized by a low scale, and may not be
calculable. The vertex and spectator corrections are handled in
the way stated above. The same formalism is then applied to the
other exclusive $B$ meson decays into charmonia in Sec.~III.
Section IV is the conclusion. The Appendix collects the
expressions of the involved meson distribution amplitudes and of
the factorization formulas for all the spectator amplitudes. For a
review of the studies of semi-inclusive $B$ meson decays into
charmonia, refer to \cite{yellow}.



\section{THE $B\to J/\psi K^{(*)}$ DECAYS}

We write the $B$ ($J/\psi$) meson momentum $P_1$ ($P_2$) in the
light-cone coordinates as
\begin{eqnarray}
P_1=\frac{m_B}{\sqrt{2}}(1,1,{\bf 0}_T)\;,\;\;\;\;
P_2=\frac{m_B}{\sqrt{2}}\left(1,r_2^2, {\bf
0}_T\right)\;,\label{mom}
\end{eqnarray}
with the mass ratio $r_2=m_{J/\psi}/m_B$. The $K^{(*)}$ meson
momentum is then given by $P_3=P_1-P_2$. The polarization vectors
of the $J/\psi$ meson are parameterized as
\begin{eqnarray}
\epsilon_{2L}=\frac{1}{\sqrt{2}r_2}\left(1,-r_2^2, {\bf
0}_T\right)\;,\;\;\;\; \epsilon_{2T}=\left(0,0, {\bf
1}_T\right)\;.\label{pol}
\end{eqnarray}
The relevant effective Hamiltonian for the $B\to J/\psi K^{(*)}$
decays is
\begin{eqnarray}
H_{\rm eff}&=&\frac{G_F}{\sqrt
2}\left\{V_{cb}V_{cs}^{*}[C_1(\mu)O_1+
C_2(\mu)O_2]-V_{tb}V_{ts}^{*}\sum_{k=3}^{10}C_k(\mu)O_k\right\}\;,
\label{effj}
\end{eqnarray}
with the Cabibbo-Kobayashi-Maskawa (CKM) matrix elements $V$ and
the four-fermion operators,
\begin{eqnarray}
& &O_1 =
(\bar{s}_ic_j)_{V-A}(\bar{c}_jb_i)_{V-A}\;,\;\;\;\;\;\;\;\; O_2 =
(\bar{s}_ic_i)_{V-A}(\bar{c}_jb_j)_{V-A}\;,
\nonumber \\
& &O_3
=(\bar{s}_ib_i)_{V-A}\sum_{q}(\bar{q}_jq_j)_{V-A}\;,\;\;\;\; O_4
=(\bar{s}_ib_j)_{V-A}\sum_{q}(\bar{q}_jq_i)_{V-A}\;,
\nonumber \\
& &O_5
=(\bar{s}_ib_i)_{V-A}\sum_{q}(\bar{q}_jq_j)_{V+A}\;,\;\;\;\;O_6
=(\bar{s}_ib_j)_{V-A}\sum_{q}(\bar{q}_jq_i)_{V+A}\;,
\nonumber \\
& &O_7
=\frac{3}{2}(\bar{s}_ib_i)_{V-A}\sum_{q}e_q(\bar{q}_jq_j)_{V+A}\;,
\;\; O_8
=\frac{3}{2}(\bar{s}_ib_j)_{V-A}\sum_{q}e_q(\bar{q}_jq_i)_{V+A}\;,
\nonumber \\
& &O_9
=\frac{3}{2}(\bar{s}_ib_i)_{V-A}\sum_{q}e_q(\bar{q}_jq_j)_{V-A}\;,
\;\; O_{10}
=\frac{3}{2}(\bar{s}_ib_j)_{V-A}\sum_{q}e_q(\bar{q}_jq_i)_{V-A}\;,
\end{eqnarray}
$i, \ j$ being the color indices. Below we shall neglect the tiny
product $V_{ub}V_{us}^{*}$, and assume
$V_{cb}V_{cs}^{*}=-V_{tb}V_{ts}^{*}$ for convenience.

\subsection{Factorization Formulas}

The $B\to J/\psi K$ decay rate has the expression,
\begin{equation}
\Gamma=\frac{1}{32\pi}G_F^2|V_{cb}|^2|V_{cs}|^2m_B^3(1-r_2^2)^3
|{\cal A}|^2\;.
\end{equation}
Computing the factorizable contribution to the amplitude ${\cal
A}$ in PQCD, we found that its characteristic scale is around 1
GeV. With such a small hard scale, the perturbation theory may not
be reliable. Therefore, we do not attempt to calculate the
factorizable contribution, but parameterize it in FA. The
amplitude ${\cal A}$ is then written as
\begin{eqnarray}
{\cal A}&=&a_{\rm eff}(\mu)f_{J/{\psi}}F_1(m_{J/\psi}^2)+{\cal
M}^{(J/\psi K)}\;, \label{M9}
\end{eqnarray}
with the $J/\psi$ meson decay constant $f_{J/\psi}$ and the
spectator amplitude ${\cal M}^{(J/\psi K)}$. The form factor
$F_1(q^2)$ is defined via the matrix element,
\begin{eqnarray}
\langle K(P_3)|{\bar b}\gamma_\mu s|B(P_1)\rangle
=F_1(q^2)\left[(P_1+P_3)_\mu-\frac{m_B^2-m_K^2}{q^2}q_\mu\right]
+F_0(q^2)\frac{m_B^2-m_K^2}{q^2}q_\mu\;,\label{fp}
\end{eqnarray}
$q=P_1-P_3$ being the momentum transfer, and $m_K$ the kaon mass.
The effective Wilson coefficient $a_{\rm eff}$ sums the
contributions from both the tree and penguin operators in
Eq.~(\ref{effj}):
\begin{eqnarray}
a_{\rm eff}(\mu)=a_2(\mu)+a_3(\mu)+a_5(\mu)\;,\label{a}
\end{eqnarray}
with
\begin{eqnarray}
a_2&=&C_1+\frac{C_2}{N_c}\;,\nonumber\\
a_3&=&C_3+\frac{C_4}{N_c}+\frac{3}{2}e_{c}\left( C_{9}
+\frac{C_{10}}{N_{c}}\right)\;,\nonumber\\
a_5&=&C_5+\frac{C_6}{N_c}+\frac{3}{2}e_{c} \left( C_{7}
+\frac{C_{8}}{N_{c}}\right) \;,\label{aeff}
\end{eqnarray}
and the charm quark charge $e_c=2/3$.

After parameterizing the factorizable contribution in FA, the soft
dynamics is absorbed into the form factor $F_1$. The remaining
piece, i.e., the Wilson coefficient, is dominated by hard dynamics
of $O(m_b)$, $m_b$ being the $b$ quark mass. We shall set the
renormalization scale to $\mu=m_b\approx 4.4 $ GeV, which
corresponds to $a_{\rm eff}(m_b)\approx 0.1$. The spectator
amplitude ${\cal M}^{(J/\psi)}$ is characterized by a scale of
$O(\sqrt{\bar\Lambda m_b})$ with $\bar\Lambda$ being a hadronic
scale
\cite{KKLL,NL2,CKL}:
the hard gluons in Figs.~\ref{fvs}(e) and \ref{fvs}(f), carrying
the difference between the momenta of the soft spectator in the
$B$ meson and of the energetic spectator in the kaon, are
off-shell by $O(\sqrt{\bar\Lambda m_b})$.
Strictly speaking, once a $B$ meson transition form factor is
treated as a soft object, the vertex corrections in
Figs.~\ref{fvs}(a)-\ref{fvs}(d) should be added, which become of
the same order as the spectator amplitude. This is the counting
rules of QCDF. The PQCD formalism for the vertex corrections
requires the inclusion of the transverse momenta $k_T$ of the
charm quarks, because of the end-point singularities in some
modes. The $k_T$-dependent charmonium wave functions are then the
necessary inputs, which, however, have been determined neither
theoretically nor experimentally. Therefore, the vertex
corrections appear as a theoretical uncertainty eventually, which
can be covered by a variation of the scale $\mu$ between $0.5 m_b$
and $1.5 m_b$ as shown below.

In terms of the notation in \cite{TLS}, we decompose the nonlocal
matrix elements associated with longitudinally and transversely
polarized $J/\psi$ mesons into
\begin{eqnarray}
\langle J/\psi(P,\epsilon_L)|\bar c(z)_jc(0)_l|0\rangle
&=&\frac{1}{\sqrt{2N_c}}\int_0^1 dx e^{ixP\cdot z}
\bigg\{m_{J/\psi}[\slashs \epsilon_L]_{lj}\Psi^L(x)+[\slashs
\epsilon_L\slashs P]_{lj} \Psi^{t}(x)
\bigg\}\;,
\label{lpf}\\
\langle J/\psi(P,\epsilon_T)|\bar c(z)_jc(0)_l|0\rangle
&=&\frac{1}{\sqrt{2N_c}}\int_0^1 dx e^{ixP\cdot z}
\bigg\{m_{J/\psi}[\slashs \epsilon_T]_{lj}\Psi^V(x)+
[\slashs\epsilon_T\slashs P]_{lj}\Psi^T(x)
\bigg\}\;, \label{spf}
\end{eqnarray}
respectively, which define the twist-2 distribution amplitudes
$\Psi^L$ and $\Psi^T$, and the twist-3 distribution amplitudes
$\Psi^t$ and $\Psi^V$ with the $c$ quark carrying the fractional
momentum $xP$. It is confusing that both the structures
$\slashs\epsilon_L$ and $\slashs\epsilon_L\slashs P$ were regarded
as being leading-twist in the analysis of the $B\to J/\psi K$
decays \cite{chay,Cheng}. In fact, it is
$\slashs\epsilon_{T}\slashs P$ that is leading-twist, which
contributes only to the $B\to J/\psi K^*$ decays.

\begin{figure}[t!]
\begin{center}
\epsfig{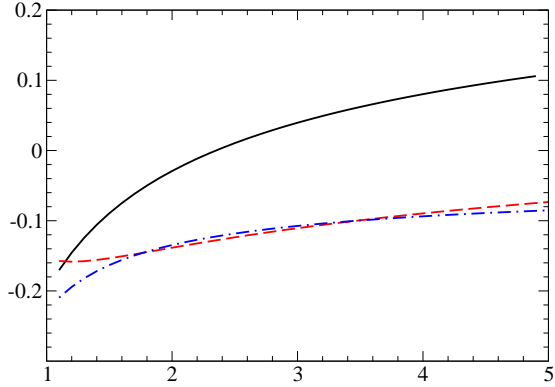}
\end{center}
\caption{Dependence of $a_{\rm eff}$ on the renormalization scale
$\mu$. The unit for the horizontal axis is GeV. The solid, dashed
and dash-dotted lines stand for $a_{\rm eff}$ without the vertex
corrections, the real part of $a_{\rm eff}$ with the vertex
corrections, and the imaginary part of $a_{\rm eff}$ with the
vertex corrections, respectively.} \label{rg}
\end{figure}

The vertex corrections to the $B\to J/\psi K$ decays, denoted as
$f_I$ in QCDF, have been calculated in the NDR scheme
\cite{chay,Cheng}. Their effect can be combined into the Wilson
coefficients associated with the factorizable contributions:
\begin{eqnarray}
a_2&=&C_1+\frac{C_2}{N_c}+\frac{\alpha_s}{4\pi}\frac{C_F}{N_c}C_2
\left(-18+12\ln\frac{m_b}{\mu}+f_I\right)\;,\nonumber\\
a_3&=&C_3+\frac{C_4}{N_c}+\frac{\alpha_s}{4\pi}\frac{C_F}{N_c}C_4
\left(-18+12\ln\frac{m_b}{\mu}+f_I\right)\;,\nonumber\\
a_5&=&C_5+\frac{C_6}{N_c}+\frac{\alpha_s}{4\pi}\frac{C_F}{N_c}C_6
\left(6-12\ln\frac{m_b}{\mu}-f_I\right)\;,\label{a235}
\end{eqnarray}
with the function,
\begin{eqnarray}
f_I=\frac{2\sqrt{2N_c}}{f_{J/\psi}}\int
dx_3\Psi^L(x_2)\left[\frac{3(1-2x_2)}{1-x_2}\ln x_2-3\pi
i+3\ln(1-r_2^2)+\frac{2r_2^2(1-x_2)}{1-r_2^2 x_2}\right]\;,
\end{eqnarray}
where those terms proportional to $r_2^4$ have been neglected. We
have also neglected the contributions from the electroweak penguin
operators for simplicity. As shown in Fig.~\ref{rg}, including the
vertex corrections leads to a larger $|a_{\rm eff}(m_b)|\approx
0.12$, which reproduces Eq.~(\ref{eq:BRnf}) roughly. We have
confirmed that
the variation of
$\mu$ between $0.5m_b$ and $1.5m_b$ is wide enough for taking into
account the effect of the vertex corrections.

The definitions of the $B$ meson wave function $\Phi_B$ and of the
kaon distribution amplitudes $\Phi_K$ are referred to
\cite{CKL,TLS}. It is then straightforward to derive the spectator
amplitude,
\begin{eqnarray}
{\cal M}^{(J/\psi K)}&=&{\cal M}_{4}^{(J/\psi K)}+{\cal
M}_{6}^{(J/\psi K)}\;,\label{mjk}
\end{eqnarray}
where the amplitudes ${\cal M}_{4}^{(J/\psi K)}$ and ${\cal
M}_{6}^{(J/\psi K)}$ result from the $(V-A)(V-A)$ and $(V-A)(V+A)$
operators in Eq.~(\ref{effj}), respectively. Their factorization
formulas are given by
\begin{eqnarray}
{\cal M}_{4}^{(J/\psi K)} &=&16\pi C_{F}\sqrt{2N_{c}}%
\int_{0}^{1}[dx]\int_{0}^{\infty }b_{1}db_{1} \Phi
_{B}(x_{1},b_{1})
\nonumber \\
&& \times \Big\{ \Big[ (1-2r^{2}_{2})(1-x_{2}) \Phi _{K} ( x_{3}
)\Psi^{L}(x_{2})
+\frac{1}{2} r^{2}_{2} \Phi_{K}(x_{3})\Psi^{t }(x_{2}) \nonumber \\
&&- r_{K} (1-r^2_2)x_3 \Phi^{p}_{K}(x_3) \Psi _{L}(x_{2})+ r_{K}
\left( 2r^{2}_{2}(1-x_2)+(1-r^2_2)x_3 \right)
\Phi^{\sigma}_{K}(x_3) \Psi^{L}(x_{2})
  \Big]\nonumber \\
&&\times E_{4}(t_d^{(1)})h_d^{(1)}(x_1,x_2,x_3,b_1)
\nonumber \\%
&& - \Big[ (x_2+(1-2r^2_{2})x_{3})\Phi _{K} ( x_{3}
)\Psi^{L}(x_{2})+r^{2}_{2}(2r_{K}\Phi^{\sigma}_{K}(x_{3})
-\frac{1}{2}\Phi_{K}(x_{3}))\Psi^{t}(x_{2})
 \nonumber \\
&& -r_{K} (1-r^2_2)x_3 \Phi^{p}_{K}(x_3) \Psi _{L}(x_{2})- r_{K}
\left( 2r^{2}_{2}x_2+(1-r^2_2)x_3 \right) \Phi^{\sigma}_{K}(x_3)
\Psi^{L}(x_{2})\Big]\nonumber \\
&& \times E_{4}(t^{(2)}_d)
h_d^{(2)}(x_1,x_2,x_3,b_1)\;,\label{psi4}\\
{\cal M}_{6}^{(J/\psi K)} &=&16\pi C_{F}\sqrt{2N_{c}}%
\int_{0}^{1}[dx]\int_{0}^{\infty }b_{1}db_{1} \Phi
_{B}(x_{1},b_{1})
\nonumber \\
&& \times \Big\{  \Big[ (1-x_2+(1-2r^2_{2})x_{3})\Phi _{K} ( x_{3}
)\Psi^{L}(x_{2})+
r^{2}_{2}(2r_{K}\Phi^{\sigma}_{K}(x_{3})-\frac{1}{2}\Phi_{K}(x_{3}))\Psi^{t}(x_{2})
 \nonumber \\
&& -r_{K} (1-r^2_2)x_3 \Phi^{p}_{K}(x_3) \Psi^{L}(x_{2})- r_{K}
\left( 2r^{2}_{2}(1-x_2)+(1-r^2_2)x_3 \right)
\Phi^{\sigma}_{K}(x_3)
\Psi^{L}(x_{2})\Big]\nonumber \\
&&\times E_{6}(t_d^{(1)})h_d^{(1)}(x_1,x_2,x_3,b_1)
\nonumber \\%
&& - \Big[ (1-2r^{2}_{2})x_{2} \Phi _{K} ( x_{3} )\Psi^{L}(x_{2})
+ \frac{1}{2} r^{2}_{2} \Phi_{K}(x_{3})\Psi^{t }(x_{2}) \nonumber \\
&&- r_{K} (1-r^2_2)x_3 \Phi^{p}_{K}(x_3) \Psi^{L}(x_{2})+ r_{K}
\left( 2r^{2}_{2}x_2+(1-r^2_2)x_3 \right) \Phi^{\sigma}_{K}(x_3)
\Psi^{L}(x_{2})
  \Big]\nonumber \\
&& \times E_{6}(t^{(2)}_d) h_d^{(2)}(x_1,x_2,x_3,b_1)
\Big\}\;,\label{psi6}
\end{eqnarray}
with the color factor $C_F=4/3$, the number of colors $N_c=3$, the
symbol $[dx]\equiv dx_1 dx_2 dx_3$ and the mass ratio
$r_K=m_0^K/m_B$, $m_0^K$ being the chiral scale associated with
the kaon.

The evolution factors are written as
\begin{eqnarray}
E_{i}(t) &=&\alpha _{s}(t) a_{i}^{\prime}(t)S(t)|_{b_{3}=b_{1}}\;,
\end{eqnarray}
with the Wilson coefficients,
\begin{eqnarray}
a_{4}^{\prime } &=&\frac{C_{2}}{N_{c}}+\frac{1}{N_{c}} \left(
C_{4}+\frac{3}{2}e_{c}C_{10}\right)\;, \nonumber\\
a_{6}^{\prime} &=&\frac{1}{N_{c}}\left(
C_{6}+\frac{3}{2}e_{c}C_{8}\right)\;.
\end{eqnarray}
The Sudakov exponent is given by
\begin{eqnarray}
S(t)&=&S_B(t)+S_K(t)\;,\nonumber\\
S_{B}(t)&=&\exp\left[-s(x_{1}P_{1}^{+},b_{1})
-\frac{5}{3}\int_{1/b_{1}}^{t}\frac{d{\bar{\mu}}} {\bar{\mu}}
\gamma (\alpha _{s}({\bar{\mu}}))\right]\;,
\label{sb} \\
S_{K}(t)&=&\exp\left[-s(x_{3}P_{3}^{-},b_{3})
-s((1-x_{3})P_{3}^{-},b_{3})
-2\int_{1/b_{3}}^{t}\frac{d{\bar{\mu}}}{\bar{\mu}} \gamma
(\alpha_{s}({\bar{\mu}}))\right]\;, \label{sbk}
\end{eqnarray}
with the quark anomalous dimension $\gamma=-\alpha_s/\pi$. Note
that the coefficient $5/3$ of the quark anomalous dimension in
Eq.~(\ref{sb}) differs from 2 in Eq.~(\ref{sbk}). The reason is
that the rescaled heavy-quark field adopted in the definition of
the heavy-meson wave function has a self-energy correction
different from that of the full heavy-quark field \cite{LL04}. The
variables $b_{1}$ and $b_{3}$, conjugate to the parton transverse
momenta $k_{1T}$ and $k_{3T}$, represent the transverse extents of
the $B$ and $K$ mesons, respectively. The explicit expression of
the exponent $s$ can be found in \cite{CS,BS,LS}. The above
Sudakov exponentials decrease fast in the large $b$ region, such
that the $B\to J/\psi K$ spectator amplitudes remain sufficiently
perturbative in the end-point region of the momentum fractions.

The hard functions $h_d^{(j)}$, $j=1$ and 2, are
\begin{eqnarray}
h^{(j)}_d&=& \frac{1}{D-D_j} \left( \begin{array}{cc}
 K_{0}(\sqrt{D_{j}}m_Bb_{1})-K_{0}(\sqrt{D}m_Bb_{1}) &  \mbox{for $D_{j} \geq 0$}  \\
 \frac{i\pi}{2} H_{0}^{(1)}\left(\sqrt{|D_{j}|}m_Bb_{1}\right)-K_{0}(\sqrt{D}m_Bb_{1})
   & \mbox{for $D_{j} < 0$}
  \end{array} \right)\;,
\label{hjd}
\end{eqnarray}
with the arguments,
\begin{eqnarray}
D& =& x_1x_3(1-r_2^{2})\;,  \\
D_1& =&(1-x_2)x_1r_2^{2}+(x_1+x_2-1)x_3(1-r_2^{2})
+\left[\frac{1}{4}-(1-x_2)^2\right]r_2^2\;,  \\
D_2& =& x_1x_2r_2^{2}+(x_1-x_2)x_3(1-r_2^{2})
+\left(\frac{1}{4}-x_2^2\right)r_2^2\;.
\end{eqnarray}
The hard scales $t$ are chosen as
\begin{eqnarray}
t^{(j)}={\rm max}(\sqrt{D}m_B,\sqrt{|D_j|}m_B,1/b_1)\;.
\end{eqnarray}
Without the transverse momenta $k_T$, the terms containing
$2r_Kr_2^2\Phi^{\sigma}_{K}\Psi^{t}$ in Eqs.~(\ref{psi4}) and
(\ref{psi6}) are logarithimically divergent due to the end-point
singularities from $x_3\to 0$ \cite{Cheng}. Since the end-point
singularities appear only at the power of $r_Kr_2^2$, instead of
at leading power, the spectator contributions are expected to be
evaluated more reliably than the factorizable contributions in
PQCD.

For the $B\to J/\psi K^*$ modes, the parametrization of the
kinematic variables in Eqs.~(\ref{mom}) and (\ref{pol}) still
apply. We further define the polarization vectors $\epsilon_3$ for
the $K^*$ meson, which is orthogonal to the $K^*$ meson momentum
$P_3$. In this case the $K^*$ meson mass should be kept in the
intermediate stage of the calculation, and then dropped at the end
according to the power counting rules in \cite{CKL}. The decay
rate is given by
\begin{equation}
\Gamma=\frac{1}{32\pi}G_F^2|V_{cb}|^2|V_{cs}|^2m_B^3(1-r_2^2)^3
\sum_\sigma {\cal A}^{(\sigma)\dagger }{\cal A^{(\sigma)}}\;.
\end{equation}
The amplitudes for different final helicity states are expressed
as
\begin{eqnarray}
{\cal A^{(\sigma)}}&=&- \Bigg\{a_{\rm eff}(\mu)f_{J/\psi}
F(m_{J/\psi}^2)+{\cal M}^{(J/\psi K^*)}_{L},\nonumber\\
& & \epsilon^{*}_{2T}\cdot\epsilon^{*}_{3T}\left[r_2a_{\rm
eff}(\mu)f_{J/\psi}A_1(m_{J/\psi}^2)+{\cal M}^{(J/\psi
K^*)}_{N}\right],
\nonumber\\
& & i\, \epsilon^{\alpha \beta\gamma \rho}
\epsilon^{*}_{2\alpha}\epsilon^{*}_{3\beta} \frac{P_{2\gamma
}P_{3\rho }}{m_B^2}\, \left[r_2a_{\rm
eff}(\mu)f_{J/\psi}V(m_{J/\psi}^2)+{\cal M}^{(J/\psi
K^*)}_{T}\right]\Bigg\}\;, \label{M1}
\end{eqnarray}
with the combination of the form factors,
\begin{eqnarray}
F(m_{J/\psi}^2)=\frac{1+r_{K^*}}{2r_{K^*}}A_1(m_{J/\psi}^2)
-\frac{1-r_2^2}{2r_{K^*}(1+r_{K^*})}A_2(m_{J/\psi}^2)\;,
\end{eqnarray}
and the mass ratio $r_{K^*}=m_{K^*}/m_B$. The first term in
Eq.~(\ref{M1}) corresponds to the configuration with both the
vector mesons being longitudinally polarized, and the second
(third) term to the two configurations with both the vector mesons
being transversely polarized in the parallel (perpendicular)
directions.

The effective Wilson coefficient $a_{\rm eff}$ is the same as in
Eq.~(\ref{a}). The $B\to K^*$ transition form factors $A_1$,
$A_2$, and $V$ are defined via the matrix elements,
\begin{eqnarray}
\langle K^*(P_3,\epsilon_3)| \bar b \gamma^\mu s |B(P_1) \rangle&
=&
 \frac{2iV(q^2)}{m_B+m_{K^*}} \epsilon^{\mu\nu\rho\sigma}
 \epsilon_{3\nu}^{\ast}  P_{3\rho} P_{1\sigma},
\label{V}\\
\langle K^*(P_3,\epsilon_3)|\bar b \gamma^\mu\gamma_5 s | B(P_1)
\rangle &=&
  2m_{K^*}A_0(q^2)\frac{\epsilon_3^\ast\cdot q}{q^2}q^\mu +
  (m_B+m_{K^*})A_1(q^2)\left(\epsilon_3^{\ast\mu}-
  \frac{\epsilon_3^\ast\cdot q}{q^2}q^\mu\right)
\nonumber\\
&& -A_2(q^2)\frac{\epsilon_3^\ast\cdot q}{m_B+m_{K^*}}
 \left(P_1^\mu+P_3^{\mu}
 -\frac{m_B^2-m_{K^*}^2}{q^2}q^\mu\right)\;.\label{a012}
\end{eqnarray}
The spectator amplitudes ${\cal M}^{(J/\psi K^*)}_{L,N,T}$ from
Figs.~\ref{fvs}(e) and \ref{fvs}(f) are written as
\begin{eqnarray}
{\cal M}^{(J/\psi K^*)}_{L,N,T}={\cal M}^{(J/\psi
K^*)}_{L4,N4,T4}+{\cal M}^{(J/\psi K^*)}_{L6,N6,T6}\;,
\end{eqnarray}
where the explicit factorization formulas for the amplitudes
${\cal M}^{(J/\psi K^*)}_{L4,N4,T4}$ and for ${\cal M}^{(J/\psi
K^*)}_{L6,N6,T6}$ are presented in the Appendix.

The helicity amplitudes are then defined as,
\begin{eqnarray}
A_{L}&=&G\left[a_{\rm eff}(\mu)f_{J/\psi}
F(m_{J/\psi}^2)+{\cal M}^{(J/\psi K^*)}_{L}\right]\;, \nonumber\\
A_{\parallel}&=&-G \sqrt{2}\left[r_2a_{\rm
eff}(\mu)f_{J/\psi}A_1(m_{J/\psi}^2)+{\cal M}^{(J/\psi
K^*)}_{N}\right]\;, \nonumber \\
A_{\perp}&=&-G r_2r_{K^*} \sqrt{2[(v_2\cdot v_3)^{2}-1]}
\left[r_2a_{\rm eff}(\mu)f_{J/\psi}V(m_{J/\psi}^2)+{\cal
M}^{(J/\psi K^*)}_{T}\right]\;, \label{ase}
\end{eqnarray}
for the longitudinal, parallel, and perpendicular polarizations,
respectively, with the velocities $v_2=P_2/m_{J/\psi}$ and
$v_3=P_3/m_{K^*}$. The normalization factor $G$ has been chosen
such that the following relation is satisfied:
\begin{eqnarray}
|A_{L}|^2+|A_{\parallel}|^2+|A_{\perp}|^2=1\;.
\end{eqnarray}
In our convention the relative phase $\delta_\parallel$
($\delta_\perp$) between $A_{\parallel}$ ($A_{\perp}$) and $A_{L}$
takes the value of $\pi$ in the heavy-quark limit, ie., as $a_{\rm
eff}$ is real and ${\cal M}^{(J/\psi K^*)}_{L,N,T}$ vanish. We
define the polarization fractions $f_L=|A_{L}|^2$,
$f_\parallel=|A_{\parallel}|^2$, and $f_\perp=|A_{\perp}|^2$. The
$B\to J/\psi K^*$ polarizations, significantly affected by the
nonfactorizable contributions, were not discussed in \cite{LM04}.

\subsection{Numerical Analysis}

To perform the numerical analysis, we need the information of the
involved meson distribution amplitudes. The $B$ meson wave
function $\Phi_B$ and the $K^{(*)}$ meson distribution amplitudes
have been studied intensively, whose expressions are collected in
the Appendix. The charmonium distribution amplitudes attract less
attention in the literature. In \cite{MASI} they were assumed to
be identical to the corresponding light-meson distribution
amplitudes: those of the $J/\psi$ ($\eta_c$) meson are the same as
of the $\rho$ ($\pi$) meson. Then the twist-3 charmonium
distribution amplitudes do not vanish at the end points $x=0$, 1,
where the valence charm quarks become highly off-shell. Such a
functional shape seems to contradict to the intuition. More
realistic asymptotic models have been proposed recently in
\cite{BC04}, which were inferred from the non-relativistic heavy
quarkonium bound-state wave functions with the same quantum
numbers. We shall adopt these $J/\psi$ and $\eta_c$ meson
distribution amplitudes, and derive the $\chi_{c0}$ and
$\chi_{c1}$ meson distribution amplitudes following the similar
procedure in the next section.

The $J/\psi$ meson asymptotic distribution amplitudes are given by
\cite{BC04}
\begin{eqnarray}
\Psi^L(x)&=&\Psi^T(x)=9.58\frac{f_{J/\psi}}{2\sqrt{2N_c}}x(1-x)
\left[\frac{x(1-x)}{1-2.8x(1-x)}\right]^{0.7}\;,\nonumber\\
\Psi^t(x)&=&10.94\frac{f_{J/\psi}}{2\sqrt{2N_c}}(1-2x)^2
\left[\frac{x(1-x)}{1-2.8x(1-x)}\right]^{0.7}\;,\nonumber\\
\Psi^V(x)&=&1.67\frac{f_{J/\psi}}{2\sqrt{2N_c}}\left[1+(2x-1)^2\right]
\left[\frac{x(1-x)}{1-2.8x(1-x)}\right]^{0.7}\;,\label{jda}
\end{eqnarray}
in which the twist-3 ones $\Psi^{t,V}$ vanish at the end points
due to the additional factor $[x(1-x)]^{0.7}$. Compared to
\cite{YL}, we have distinguished the distribution amplitudes
associated with the longitudinally and transversely polarized
$J/\psi$ mesons, which exhibit the different asymptotic behaviors.
We do not distinguish the decay constant $f_{J/\psi}$ for the
normalization of $\Psi^L$ and $f_{J/\psi}^T$ for the normalization
of $\Psi^t$. The distribution amplitudes associated with the
structures $I$ (the identity) and $\gamma^\mu\gamma_5$ (the
pseudo-vector) diminish like $1-2m_c/m_{J/\psi}$, $m_c$ being the
$c$ quark mass. The above models have been shown to yield the
observed cross section of charmonium production in $e^+e^-$
collisions \cite{BC04}. If factorization theorem works, the same
distribution amplitudes, due to their universality, should be able
to explain the exclusive $B$ meson decays into charmonia.

\begin{table}[ht]
\begin{center}
\begin{tabular}{c|ccc} \hline
\rule[-3mm]{0mm}{8mm}   &
$F(0)$ & $a$ & $b$  \\
\hline $F_1(q^2)$ &  0.35 &  1.58 & 0.68  \\
\hline $F_0(q^2)$ &  0.35 &  0.71 & 0.04  \\
\hline $V(q^2)$ &  0.31 &  1.79 & 1.18  \\
\hline $A_0(q^2)$ &  0.31 &  1.68 & 1.08  \\
\hline $A_1(q^2)$ &  0.26 &  0.93 & 0.19  \\
\hline $A_2(q^2)$ &  0.24 &  1.63 & 0.98  \\
\hline
\end{tabular}
\end{center}
\caption{Parameters for the $B\to K^{(*)}$ transition form
factors.}\label{tab:1}
\end{table}

\begin{table}[ht]
\begin{center}
\begin{tabular}{c|cc} \hline
\rule[-3mm]{0mm}{8mm} $X(J^{PC})$ &
$m_{X}$[GeV]\protect{\cite{PDG}} & $f_{X}$[MeV]  \\
\hline $\eta_c(0^{-+})$ &  2.980 &  $420 \pm 50$
\protect{\cite{hwangkim}}   \\
\hline $J/\psi(1^{--})$ &  3.097 & 405 $\pm$ 14 \protect{\cite{Melic02}}   \\
\hline $\chi_{c0}(0^{++})$ &  3.415 &  360
\protect{\cite{novikov}}
  \\
\hline $\chi_{c1}(1^{++})$ &  3.511 & 335 \protect{\cite{novikov}}    \\
\hline
\end{tabular}
\end{center}
\caption{Decay constants and masses of various charmonium
states.}\label{tab:proper}
\end{table}

For the $B\to K^{(*)}$ transition form factors, we employ the
models derived from the light-front QCD \cite{CCH}, which have
been parameterized as
\begin{eqnarray}
F(q^2)=\frac{F(0)}{1-a(q^2/m_B^2)+b(q^2/m_B^2)^2}\;,\label{fpa}
\end{eqnarray}
with the constants $F(0)$, $a$, and $b$ being listed in
Table~\ref{tab:1}. Using the inputs in Table~\ref{tab:proper}
\cite{Melic04} for the various charmonium states, the CKM matrix
elements $V_{cb}=0.040$ and $V_{cs}=0.996$, the quark mass $m_t
=174.3 \; {\rm GeV}$ for the Wilson coefficients, the lifetimes
$\tau_{B^0}= 1.56\times 10^{-12}\;{\rm sec}$ and $\tau_{B^\pm}\,
=\, 1.67\times 10^{-12} \;{\rm sec}$, and the Fermi constant
$G_F\, =\, 1.16639\times 10^{-5}\;{\rm GeV}^{-2}$, we derive the
branching ratios,
\begin{eqnarray}
& &B(B^+ \to J/\psi K^+) = (9.20^{+6.03}_{-7.99}) \times 10^{-4}\;, \nonumber \\
& &B(B^0 \to J/\psi K^0) = (8.60^{+5.63}_{-7.47}) \times 10^{-4}\;, \nonumber \\
& &B(B^+ \to J/\psi K^{*+}) =(9.95^{+5.2}_{-7.16})  \times 10^{-4}\;, \nonumber \\
& &B(B^0 \to J/\psi K^{*0}) = (9.30^{+4.86}_{-6.69}) \times
10^{-4}\;.
\end{eqnarray}
We have also derived the the polarization fractions and the
relative phases among the helicity amplitudes for the $B \to
J/\psi K^{*}$ decays,
\begin{eqnarray}
& &f_{L}=0.73^{+0.15}_{-0.05}, \ \ \
f_{\parallel}=0.17^{+0.04}_{-0.10}, \ \ \
f_{\perp}=0.10^{+0.01}_{-0.05}\;,\nonumber\\
 & &\delta_\parallel=2.57^{+0.12}_{-0.93}   \;,\;\; \delta_\perp=2.53^{+0.13}_{-0.90}\;.
\end{eqnarray}
Both the $B^+ \to J/\psi K^{*0+}$ and $B^0 \to J/\psi K^{*0}$
modes have the same polarization fractions, since they differ only
in the lifetimes in our formalism. The theoretical uncertainties
come from the variation of the renormalization scale $\mu$.
We have checked the sensitivity of our results to
the choice of the $b$ quark mass: $m_b=4.8$ GeV increases the
branching ratio $B(B^0 \to J/\psi K^{0})$ only by 10\%.

The recent Babar measurement gave \cite{Babar-hep04}
\begin{eqnarray}
& &B(B^+ \to J/\psi K^{*+}) = (14.54\pm 0.47\pm 0.97)\times 10^{-4}\;, \nonumber \\
& &B(B^0 \to J/\psi K^{*0}) = (13.09\pm 0.26\pm 0.77)\times
10^{-4}\;\label{bpks}
\end{eqnarray}
and the Belle measurement gave \cite{Paoti}
\begin{eqnarray}
 & &f_{L}=0.585\pm 0.012\pm 0.009, \ \ \
f_{\parallel}=0.233\pm 0.013\pm 0.008, \ \ \
f_{\perp}=0.181\pm 0.012\pm 0.008\;,\nonumber\\
 & &\delta_\parallel=2.888\pm 0.090\pm 0.008 \;, \;\;\delta_\perp=2.903\pm 0.064\pm 0.010\;.
\end{eqnarray}
It is found that the consistency between the theoretical and
experimental values for the $B\to J/\psi K^{(*)}$ branching ratios
is reasonable. The predicted longitudinal polarization fraction
$f_L$ for the $B\to J/\psi K^*$ is a bit larger, and the predicted
relative phases are a bit smaller than the data. However, we point
out that the polarization fractions and the relative phases could
be modified by adding the higher Gegenbauer terms to the $J/\psi$
meson distribution amplitudes, and that the consistency will be
improved. This fine tuning will not be performed here. Roughly
speaking, it is very promising to understand the $B\to J/\psi
K^{(*)}$ data in the PQCD framework.

Another remark is as follows. Most of the $J/\psi$ meson
distribution amplitudes, except $\Psi^t$, exhibit maxima at
$x=1/2$, such that the valence charm quark, carrying the invariant
mass $x^2P_2^2\approx m_c^2$, is almost on-shell. However,
$\Psi^t$ has two humps with a dip at $x=1/2$. We argue that the
models in Eq.~(\ref{jda}) are the asymptotic ones. Including the
higher Gegenbauer terms, such as those for a light vector meson,
\begin{eqnarray}
\Psi^t(x)&=&10.94\frac{f_{J/\psi}}{2\sqrt{2N_c}}\left\{
(1-2x)^2+a_1^{J/\psi}(2x-1)^2[5(2x-1)^2-3]\right.
\nonumber \\
& &\left.+a_2^{J/\psi}[3-30(2x-1)^2+35(2x-1)^4]\right\}
\left[\frac{x(1-x)}{1-2.8x(1-x)}\right]^{0.70}\;,
\end{eqnarray}
$\Psi^t$ possess three humps with the major one located at
$x=1/2$. Adopting this parametrization, the Gegenbauer
coefficients $a_1^{J/\psi}$ and $a_2^{J/\psi}$ should be
determined from the data of charmonium production in $e^+e^-$
annihilation, before it is applied to the exclusive $B$ meson
decays into charmonia.


\section{$B\to (\chi_{c0}, \chi_{c1}, \eta_c) K^{(*)}$ DECAYS}

Before calculating the spectator contributions to other similar
modes, we clarify a confusing statement in the literature. It has
been claimed \cite{Chao2} that the end-point singularities from
$x_3\to 0$ cancel between the spectator diagrams
Figs.~\ref{fvs}(e) and \ref{fvs}(f) in the QCDF formalism for the
$B\to J/\psi K$ decays, but do not for the $B\to \chi_{c1}K$
decays. Our opinion is that the end-point singularities should
appear in both modes at the power of $r_2^2$, if no approximation
is made. The singularities cancel in \cite{Cheng} because of the
relation,
\begin{eqnarray}
\frac{f_{J/\psi}^T m_c}{f_{J/\psi} m_{J/\psi}}=2
x_2^2\;,\label{onsh}
\end{eqnarray}
where the decay constant $f_{J/\psi}^T$ is associated with the
normalization of the twist-3 distribution amplitude $\Psi^t$. This
relation arises from the exact on-shell condition of the valence
charm quarks. The singularities were present in \cite{Chao2},
since the authors, considering only the leading-twist $\chi_{c1}$
meson distribution amplitude, got no chance to apply
Eq.~(\ref{onsh}). We agree on the comment made in \cite{Cheng}: it
is not clear why the singularities cancel in \cite{chay} for the
$B\to J/\psi K$ decays, viewing that they treated the ratio on the
left-hand side of Eq.~(\ref{onsh}) as a constant. The exact
on-shell condition is indeed not necessary, because a deviation
from mass shell, as long as being power-suppressed, is allowed.
Furthermore, the momentum fraction $x_2$, running between 0 and 1,
is hardly related to the left-hand side. Once Eq.~(\ref{onsh}) is
not postulated, the end-point singularities exist in the terms
proportional to $r_2^2$ for the $B\to J/\psi K$ decays \cite{YL},
and also for the $B\to \chi_{c1} K$ decays. Note that these
singularities disappear in the PQCD approach based on $k_T$
factorization theorem.

The PQCD analysis of the $B\to\chi_{c1}K^{(*)}$ decays is similar
to that of $B\to J/\psi K^{(*)}$. To calculate the nonfactorizable
spectator amplitudes, we consider the $\chi_{c1}$ meson
distribution amplitudes defined via the nonlocal matrix elements
associated with the longitudinal and transverse polarizations,
\begin{eqnarray}
\langle \chi_{c1}(P,\epsilon_L)|\bar c(z)_jc(0)_l|0\rangle
&=&\frac{1}{\sqrt{2N_c}}\int_0^1 dx e^{ixP\cdot z}
\bigg\{m_{\chi_{c1}}[\gamma_5\slashs
\epsilon_L]_{lj}\chi_{1}^{L}(x)+[\gamma_5\slashs \epsilon_L\slashs
P]_{lj} \chi_{1}^{t}(x) \bigg\},
\label{lpf1}\\
\langle \chi_{c1}(P,\epsilon_T)|\bar c(z)_jc(0)_l|0\rangle
&=&\frac{1}{\sqrt{2N_c}}\int_0^1 dx e^{ixP\cdot z}
\bigg\{m_{\chi_{c1}}[\gamma_5\slashs \epsilon_T]_{lj}\chi_1^V(x)+
[\gamma_5\slashs\epsilon_T\slashs P]_{lj}\chi_1^T(x) \bigg\},
\label{spf1}
\end{eqnarray}
respectively, with the $\chi_{c1}$ meson mass $m_{\chi_{c1}}$. The
asymptotic models for the twist-2 distribution amplitudes
$\chi_1^L(x)$ and $\chi_1^T(x)$, and for the twist-3 distribution
amplitudes $\chi_1^t(x)$ and $\chi_1^V(x)$ will be derived
following the prescription in \cite{BC04}: they can be extracted
from the $n=2$, $l=1$ Schrodinger states for a Coulomb potential,
whose radial dependence is given by
\begin{eqnarray}
\Psi_{\rm Sch}(r)\propto r\exp(-q_B r)\;,
\end{eqnarray}
$q_B$ being the Bohr momentum. Fourier transformation of the above
solution leads to
\begin{eqnarray}
\Psi_{\rm Sch}(k)\propto
\frac{k^2-3q_B^2}{(k^2+q_B^2)^3}\;,\label{sch}
\end{eqnarray}
with $k^2=|{\bf k}|^2$. Employing the substitution \cite{BC04},
\begin{eqnarray}
{\bf k}_T\to {\bf k}_T\;,\;\;\;k_z\to (x-\bar
x)\frac{m_D}{2}\;,\;\;\; m_D^2=\frac{m_c^2+k_T^2}{x\bar x}\;,
\end{eqnarray}
$x$ ($\bar x\equiv 1-x$) being the $c$ ($\bar c$) quark momentum
fraction, we obtain the heavy quarkonium distribution amplitude,
\begin{eqnarray}
\Phi(x)\sim \int d^2k_T\Psi_{\rm Sch}(x,k_T)\propto x\bar
x\left\{\frac{x \bar x[1-4x\bar x(1+v^2)]}{[1-4x\bar
x(1-v^2)]^2}\right\}\;,\label{x1}
\end{eqnarray}
with $v^2=q_B^2/m_c^2$.

To impose a fractional power $1-v^2$ to the factor in the curved
brackets \cite{BC04}, we neglect the $v^2$ term in the numerator.
Otherwise, the numerator may become negative for
\begin{eqnarray}
1-\sqrt{1-\frac{1}{1+v^2}}<2x<1+\sqrt{1-\frac{1}{1+v^2}}\;,
\end{eqnarray}
and the analyticity of the distribution amplitude will be lost. We
then propose the $\chi_{c1}$ meson distribution amplitudes
inferred from Eq.~(\ref{x1}),
\begin{eqnarray}
\chi_1(x)\propto \Phi^{\rm asy}(x)\left\{\frac{x \bar x(1-4x\bar
x)}{[1-4x\bar x(1-v^2)]^2}\right\}^{1-v^2}\;,\label{int}
\end{eqnarray}
with $\Phi^{\rm asy}(x)$ being set to the asymptotic models of the
corresponding twists for light vector mesons. It is now clear that
after imposing the fractional power $1-v^2$, Eq.~(\ref{int})
approaches the light meson distribution amplitudes as $v^2\to 1$,
and the heavy quarkonium distribution amplitudes the same as
Eq.~(\ref{x1}) as $v^2\to 0$ \cite{BC04}. Fixing the
normalization, we derive
\begin{eqnarray}
\chi_1^L(x)&=&\chi_1^T(x)=27.46\frac{f_{\chi_{c1}}}{2\sqrt{2N_c}}x(1-x)
\left\{\frac{x(1-x)[1-4x(1-x)]}{[1-2.8x(1-x)]^2}\right\}^{0.7}\;,\nonumber\\
\chi_1^t(x)&=&15.17\frac{f_{\chi_{c1}}}{2\sqrt{2N_c}}(1-2x)^2
\left\{\frac{x(1-x)[1-4x(1-x)]}{[1-2.8x(1-x)]^2}\right\}^{0.7}\;,\nonumber\\
\chi_1^V(x)&=&3.60\frac{f_{\chi_{c1}}}{2\sqrt{2N_c}}\left[1+(2x-1)^2\right]
\left\{\frac{x(1-x)[1-4x(1-x)]}{[1-2.8x(1-x)]^2}\right\}^{0.7}\;,\label{chda}
\end{eqnarray}
where the same decay constant $f_{\chi_{c1}}$ has been assumed for
the longitudinally and transversely polarized $\chi_{c1}$ meson.
Note that the dip at $x=1/2$ is a consequence of the $P$-wave
Schrodinger wave functions.

The $B\to\chi_{c1}K^{(*)}$ decay amplitudes are written as
\begin{eqnarray}
{\cal A}&=&a_{\rm eff}(\mu)f_{\chi_{c1}}F_1(m_{\chi_{c1}}^2)+{\cal
M}^{(\chi_{c1} K)}\;, \label{M8}\\
{\cal A^{(\sigma)}}&=&- \Bigg\{a_{\rm eff}(\mu)f_{\chi_{c1}}
F(m_{\chi_{c1}}^2)+{\cal M}^{(\chi_{c1} K^*)}_{L},\nonumber\\
& & \epsilon^{*}_{2T}\cdot\epsilon^{*}_{3T}\left[a_{\rm
eff}(\mu)r_2f_{\chi_{c1}}A_1(m_{\chi_{c1}}^2)+{\cal M}^{(\chi_{c1}
K^*)}_{N}\right],
\nonumber\\
& & i\, \epsilon^{\alpha \beta\gamma \rho}
\epsilon^{*}_{2\alpha}\epsilon^{*}_{3\beta} \frac{P_{2\gamma
}P_{3\rho }}{m_B^2}\,
\left[r_2f_{\chi_{c1}}V(m_{\chi_{c1}}^2)+{\cal M}^{(\chi_{c1}
K^*)}_{T}\right]\Bigg\}\;, \label{M7}
\end{eqnarray}
with the mass ratio $r_2=m_{\chi_{c1}}/m_B$, and the effective
Wilson coefficient,
\begin{eqnarray}
a_{\rm eff}(\mu)=a_2(\mu)+a_3(\mu)-a_5(\mu)\;.\label{b}
\end{eqnarray}
Similarly, we shall vary the renormalization scale $\mu$ in the
range between $0.5m_b$ and $1.5 m_b$, which covers the effect of
the vertex corrections. As in the case of $B\to J/\psi K$, the
infrared divergences in the vertex corrections cancel in the
$B\to\chi_{c1} K$ decays: the hard kernel of the vertex
corrections is odd for the structure $\gamma_5\slashs\epsilon$
under the exchange of $x$ and $1-x$, while the corresponding
distribution amplitude is even. The hard kernel vanishes for the
structure $\gamma_5\slashs\epsilon\slashs P$. Express the
spectator amplitudes ${\cal M}^{(\chi_{c1} K)}$ and ${\cal
M}^{(\chi_{c1} K^*)}_{L,N,T}$ as
\begin{eqnarray}
{\cal M}^{(\chi_{c1} K)}&=&{\cal M}^{(\chi_{c1} K)}_{4}+{\cal
M}^{(\chi_{c1} K)}_{6}\;,\nonumber\\
{\cal M}^{(\chi_{c1}K^*)}_{L,N,T}&=&{\cal M}^{(\chi_{c1}
K^*)}_{L4,N4,T4}+{\cal M}^{(\chi_{c1} K^*)}_{L6,N6,T6}\;,
\end{eqnarray}
where the factorization formulas of the amplitudes ${\cal
M}^{(\chi_{c1} K)}_{4}$, ${\cal M}^{(\chi_{c1} K)}_{6}$, ${\cal
M}^{(\chi_{c1} K^*)}_{L4,N4,T4}$ and ${\cal M}^{(\chi_{c1}
K^*)}_{L6,N6,T6}$ are shown in the Appendix. Their expressions are
similar to those of $B\to J/\psi K^{(*)}$ with some terms flipping
signs.

Using the inputs listed in Table~\ref{tab:proper} \cite{Melic04},
we get
\begin{eqnarray}
& &B(B^+ \to \chi_{c1} K^+)=(3.15^{+3.17}_{-2.61})\times 10^{-4}\;,\nonumber\\
& &B(B^0 \to \chi_{c1} K^0) = (2.94^{+2.97}_{-2.43}) \times
10^{-4}\;,
\nonumber \\
& &B(B^+ \to \chi_{c1} K^{*0+})=(2.99^{+0.40}_{-0.50})\times 10^{-3}\;, \nonumber \\
& &B(B^0 \to \chi_{c1} K^{*0})=(2.79^{+0.37}_{-0.46})
\times 10^{-3}\;, \nonumber \\
& &f_{L}=0.38^{+0.03}_{-0.08} ,\ \ \
f_{\parallel}=0.07^{+0.01}_{-0.01}, \ \ \
f_{\perp}=0.55^{+0.06}_{-0.05},\nonumber\\
& &\delta_\parallel=2.61^{+0.16}_{-0.43}   \;,\;\;
\delta_\perp=2.58^{+0.07}_{-0.03}\;.\label{chi01}
\end{eqnarray}
The errors of the above predictions arise from the variation of
the renormalization scale $\mu$ for the factorizable contribution.
For the $B\to \chi_{c1} K^*$ branching ratios, the imaginary part
of the spectator amplitude ${\cal M}^{(\chi_{c1}K^*)}_T$
dominates, such that the influence of the $\mu$ dependence becomes
mild. This is also the reason we obtain a large $f_\perp$. The
$B\to\chi_{c1} K$ branching ratios in Eq.~(\ref{chi01}) are
consistent with the recent Babar measurement \cite{Babar-hep04},
\begin{eqnarray}
B(B^+ \to \chi_{c1} K^+) &=& (5.79 \pm 0.26 \pm 0.65) \times
10^{-4}\; , \nonumber \\
B(B^0 \to \chi_{c1} K^0) &=& (4.53\pm 0.41 \pm 0.51) \times
10^{-4}\;.
\end{eqnarray}
Note that these branching ratios, with a parametrization for the
logarithmical end-point singularities, were estimated to be only
about $10^{-4}$ in QCDF \cite{Chao2}. Our predictions for the
$B\to \chi_{c1} K^*$ decays, including the branching ratios, the
polarization fractions (especially the large $f_\perp$), and the
relative phases, can be compared with data in the future.

We then discuss the $B \to \chi_{c0} K$ decays, for which the
nonfactorizable contributions dominate due to the absence of the
factorizable ones. The nonlocal matrix element associated with the
$\chi_{c0}$ meson is decomposed into,
\begin{eqnarray}
\langle \chi_{c0}(P)|\bar c(z)_jc(0)_l|0\rangle
&=&\frac{1}{\sqrt{2N_c}}\int_0^1 dx e^{ixP\cdot z} \bigg\{[\slashs
P]_{lj} \chi_0^{v}(x)+m_{\chi_{c0}}[I]_{lj}\chi_0^s(x)
\bigg\}\;, \label{lpf0}
\end{eqnarray}
which defines the twist-2 and twist-3 distribution amplitudes,
$\chi_0^{v}(x)$ and $\chi_0^{s}(x)$, respectively, $m_{\chi_{c0}}$
being the $\chi_{c0}$ meson mass. To satisfy the identity
$\langle\chi_{c0}(P)|\bar c\gamma_\mu c|0\rangle=0$,
$\chi_0^{v}(x)$ must be anti-symmetric under the exchange of $x$
and $1-x$. Following the similar ansatz of constructing the
$\chi_{c1}$ meson distribution amplitudes, we propose the
following asymptotic models,
\begin{eqnarray}
\chi_0^v(x)&=&27.46\frac{f_{\chi_{c0}}}{12\sqrt{2N_c}}(1-2x)
\left\{\frac{x(1-x)[1-4x(1-x)]}{[1-2.8x(1-x)]^2}\right\}^{0.7}\;,
\nonumber\\
\chi_0^s(x)&=&4.73\frac{f_{\chi_{c0}}}{2\sqrt{2N_c}}
\left\{\frac{x(1-x)[1-4x(1-x)]}{[1-2.8x(1-x)]^2}\right\}^{0.7}\;,
\label{cda}
\end{eqnarray}
where the decay constant $f_{\chi_{c0}}$ for the normalization of
$\chi_0^v$ and $\chi_0^s$ has been assumed to be equal.

In the QCDF analysis of the $B\to (J/\psi, \chi_{c1})K^{(*)}$
decays, the vertex corrections from Figs.~\ref
{fvs}(a)-\ref{fvs}(d) are infrared safe. However, for the $B \to
\chi_{c0} K$ decays, the hard kernels of the vertex corrections
are even for the structure $I$, and odd for $\slashs P$ under the
exchange of $x$ and $1-x$. The corresponding $\chi_{c0}$ meson
distribution amplitudes are also even for the structure $I$, and
odd for $\slashs P$. Therefore, the infrared divergences do not
cancel \cite{Chao2}. To regularize these divergences, a binding
energy $2m_c-m_{\chi_{c0}}$ and a gluon mass $m_g$ have been
introduced in \cite{PZ04} and \cite{MGC05}, respectively. The
vertex corrections to the $B\to\chi_{c0} K$ decays can also be
handled in the PQCD approach. For the reasons stated in the
Introduction, we do not attempt such a calculation here. Instead,
we shall demonstrate that the spectator contributions are
sufficient to account for the observed $B \to \chi_{c0} K$
branching ratios. The decay amplitudes are written as
\begin{eqnarray}
{\cal A}={\cal M}^{(\chi_{c0} K^{(*)})}\;,\;\;\;\; {\cal
M}^{(\chi_{c0} K^{(*)})}={\cal M}_4^{(\chi_{c0} K^{(*)})}+{\cal
M}_6^{(\chi_{c0} K^{(*)})}\;,\label{M6}
\end{eqnarray}
with the explicit factorization formulas of the amplitudes ${\cal
M}_{4,6}^{(\chi_{c0} K^{(*)})}$ being referred to the Appendix.
There are also logarithmic end-point singularities in these
amplitudes, if using QCDF, indicating that the PQCD approach based
on $k_T$ factorization theorem is more appropriate.

Adopting the inputs in Table~\ref{tab:proper}, we obtain the
branching ratios,
\begin{eqnarray}
& &B(B^+ \rightarrow \chi_{c0} K^+) =5.61\times 10^{-4}\;, \nonumber \\
& &B(B^0 \rightarrow \chi_{c0} K^0) = 5.24 \times 10^{-4}\;, \nonumber \\
& &B(B^+ \rightarrow \chi_{c0} K^{*+})=8.69\times 10^{-4}\;, \nonumber \\
& &B(B^0 \rightarrow \chi_{c0} K^{*0})= 8.12 \times 10^{-4}\;.
\end{eqnarray}
There is no theoretical uncertainty from the variation of $\mu$,
since the factorizable contributions vanish in this case.
Babar gave the upper bounds (the values) \cite{Babar-hep05},
\begin{eqnarray}
& &B(B^+ \to \chi_{c0} K^+) < 8.9\;\; (=4.4 \pm 3.3 \pm 0.7)
\times
10^{-4}\;, \nonumber\\
& &B(B^0 \to \chi_{c0} K^0) < 12.4\;\; (=5.3 \pm 5.0 \pm 0.8)
\times 10^{-4}\;, \label{chi2}
\end{eqnarray}
and Belle gave \cite{Bel0412}
\begin{eqnarray}
& &B(B^+ \to \chi_{c0} K^+)=(1.96 \pm 0.35 \pm
0.33^{+1.97}_{-0.26}) \times 10^{-4}\;,
\end{eqnarray}
where the third error comes from a model uncertainty. Hence, we
conclude that the above data are understandable. The
$B^+\to\chi_{c0}K^+$ branching ratio was found to be as small as
$4.2\times 10^{-5}$ in \cite{PZ04} (with a large theoretical
uncertainty from parameterizing the end-point singularities in
QCDF) and $(2-3)\times 10^{-4}$ for the gluon mass $m_g=0.5-0.2$
GeV in \cite{MGC05}. Despite of several discrepancies in their
factorization formulas for the vertex corrections and for the
spectator amplitudes, the difference in their numerical results is
not really essential due to the tunable parameters.

The analysis of the $B\to\eta_cK^{(*)}$ decays is similar to that
of $B\to \chi_{c0} K^{(*)}$. The nonlocal matrix element
associated with the $\chi_{c0}$ meson is decomposed into
\begin{eqnarray}
\langle \eta_{c}(P)|\bar c(z)_jc(0)_l|0\rangle
&=&\frac{1}{\sqrt{2N_c}}\int_0^1 dx e^{ixP\cdot z}
\bigg\{[\gamma_5\slashs P]_{lj}
\eta^{v}(x)+m_{\eta_{c}}[\gamma_5]_{lj}\eta^s(x) \bigg\}\;,
\label{epf0}
\end{eqnarray}
which defines the twist-2 and twist-3 $\eta_{c}$ meson
distribution amplitudes, $\eta^{v}(x)$ and $\eta^{s}(x)$,
respectively, $m_{\eta_{c}}$ being the $\eta_{c}$ meson mass. The
asymptotic models for the $\eta_{c}$ meson distribution amplitudes
have been determined in \cite{BC04}:
\begin{eqnarray}
\eta^v(x)&=&9.58\frac{f_{\eta_{c}}}{2\sqrt{2N_c}}x(1-x)
\left[\frac{x(1-x)}{1-2.8x(1-x)}\right]^{0.7}\;,
\nonumber\\
\eta^s(x)&=&1.97\frac{f_{\eta_{c}}}{2\sqrt{2N_c}}
\left[\frac{x(1-x)}{1-2.8x(1-x)}\right]^{0.7}\;,
 \label{eda}
\end{eqnarray}
where we have assumed the same the decay constant $f_{\eta_{c}}$
for the normalization of $\eta^v$ and $\eta^s$.

The decay amplitudes are written as
\begin{eqnarray}
{\cal A}=a_{\rm eff}(\mu)f_{\eta_c}F_0(m_{\eta}^2)+{\cal
M}^{(\eta_c K)}\;, \;\;\;\; {\cal A}=a_{\rm
eff}(\mu)f_{\eta_c}A_0(m_{\eta}^2)+{\cal M}^{(\eta_c K^*)}\;,
\label{M4}
\end{eqnarray}
for which the factorization formulas of the spectator amplitudes,
\begin{eqnarray}
{\cal M}^{(\eta_{c} K^{(*)})}={\cal M}_4^{(\eta_{c}
K^{(*)})}+{\cal M}_6^{(\eta_{c} K^{(*)})}\;,
\end{eqnarray}
can be found in the Appendix. The $B\to K$ transition form factor
$F_0$ and the $B\to K^*$ transition form factor $A_0$ have been
defined in Eqs.~(\ref{fp}) and (\ref{a012}), respectively. The
effective Wilson coefficient $a_{\rm eff}(\mu)$ for the
$B\to\chi_{c0}K^{(*)}$ decays is also given by Eq.~(\ref{b}).
Similarly, we shall vary the renormalization scale $\mu$ in the
range between $0.5m_b$ and $1.5 m_b$, which covers the effect of
the vertex corrections. The hard kernels of the vertex corrections
are even for the structure $\gamma_5$, and odd for
$\gamma_5\slashs P$ under the exchange of $x$ and $1-x$. The
corresponding $\eta_c$ meson distribution amplitudes are both even
for $\gamma_5$ and $\gamma_5\slashs P$. Therefore, the infrared
divergences in the vertex corrections to the $B\to\eta_{c} K$
decay, cancelling only for $\gamma_5\slashs P$, still exist.

Using the inputs in Table~\ref{tab:proper} \cite{Melic04} and the
form factor parametrization in Eq.~(\ref{fpa}), we obtain the
branching ratios,
\begin{eqnarray}
& &B(B^+ \rightarrow \eta_{c0} K^+)= (2.34^{+2.43}_{-2.11})\times 10^{-4}\;, \nonumber \\
& &B(B^0 \rightarrow \eta_{c0} K^0)= (2.19^{+2.13}_{-2.12})
\times 10^{-4}\;, \nonumber \\
& &B(B^+ \rightarrow \eta_{c} K^{*+})=(2.82^{+2.91}_{-2.76})\times 10^{-4}\;, \nonumber \\
& &B(B^0 \rightarrow \eta_{c} K^{*0})= (2.64^{+2.71}_{-2.58})
\times 10^{-4}\;.
\end{eqnarray}
Since the $\eta_c$ meson distribution amplitudes have no two
humps, the $B\to \eta_c K$ branching ratios are expected to be
smaller than the $B\to J/\psi K$ ones. However, the Belle
measurement \cite{Belle-PRL90},
\begin{eqnarray}
B(B^+ \rightarrow \eta_{c} K^+) &=& (1.25 \pm 0.14^{+
0.10}_{-0.12}\pm 0.38) \times 10^{-3}
\;, \nonumber\\
B(B^0 \rightarrow \eta_{c} K^0) &=& (1.23\pm 0.23^{+0.12}_{-0.16}
\pm 0.38) \times 10^{-3}\;, \label{eta1}
\end{eqnarray}
and the Babar measurement \cite{Babar-PRD70R},
\begin{eqnarray}
B(B^+ \rightarrow \eta_{c} K^+) &=& (1.34\pm 0.09\pm 0.13\pm 0.41)
\times 10^{-3}
\;, \nonumber\\
B(B^0 \rightarrow \eta_{c} K^0) &=& (1.18\pm 0.16\pm 0.13\pm 0.37)
\times 10^{-3}\;, \label{eta2}
\end{eqnarray}
are significantly larger than our predictions. The QCDF approach
gave the $B\to\eta_c K$ branching ratios about $(1.4-1.9)\times
10^{-4}$ for a rough estimate, which are also too small
\cite{Chao1}.


Because the $\eta_c$ meson distribution amplitudes in
Eq.~(\ref{eda}) have been shown to produce the observed cross
section for charmonium production in $e^+e^-$ annihilation
\cite{BC04}, they are supposed to explain the branching ratios of
the exclusive $B$ meson decays into charmonia according to the
universality, even though these two processes involve dramatically
different dynamics. Therefore, we conclude that the $B\to\eta_c K$
decays are the only puzzle, and demand more studies.

\section{CONCLUSION}

The exclusive $B$ meson decays into charmonia, $B\to X K^{(*)}$,
for $X=J/\psi, \chi_{c0}, \chi_{c1}$, and $\eta_c$ have been
analyzed in the QCDF approach in the literature. Among these
modes, those with $X=\chi_{c0}$ were claimed to be explainable,
and those with $X=J/\psi, \chi_{c1}$, and $\eta_c$ were not.
However, this conclusion is not solid due to the serious end-point
singularities in many QCDF decay amplitudes for the vertex
corrections and for the spectator contributions, and due to the
ad-hoc models of the charmonium distribution amplitudes. In this
paper we have investigated these decays in an improved framework:
the factorizable contributions are still treated in FA, since the
involved $B\to K^{(*)}$ transition form factors, evaluated at the
charmonium mass, may not be calculated perturbatively; the effect
of the vertex corrections was taken into account by varying the
renormalization scale $\mu$, because their estimation in the PQCD
approach needs additional nonperturbative information, i.e., the
distribution of the charm quark in its transverse degrees of
freedom inside a charmonium; the spectator contributions were
computed in PQCD, in which the end-point singularities are smeared
by the Sudakov factor associated with the $K^{(*)}$ meson. The
$J/\psi$ and $\eta_c$ meson distribution amplitudes have been
inferred from the $n=1$, $l=0$ Schrodinger state for a Coulomb
potential \cite{BC04}, which produce the measured cross section of
charmonium production in $e^+e^-$ collisions. To analyze the $B\to
(\chi_{c0},\chi_{c1})K^{(*)}$ decays, we have obtained the
$\chi_{c0},\chi_{c1}$ meson distribution amplitudes from the
$n=2$, $l=1$ Schrodinger states following the similar procedure.
That is, our models for the charmonium distribution amplitudes
have been constrained theoretically and experimentally.

Our investigation has indicated that only the $B\to \eta_c K$
decays exhibit a puzzle, whose branching ratios are significantly
smaller than the observed values. The data of the $B\to (J/\psi,
\chi_{c0},\chi_{c1})K$, $J/\psi K^*$ decays, including the
branching ratios, the polarization fractions, and the relative
phases among the various helicity amplitudes, are all
understandable within theoretical uncertainty. If QCD
factorization theorem works, the $\eta_c$ meson distribution
amplitudes, explaining the charmonium production data, are
supposed to explain the exclusive $B$ meson decays into charmonia.
Hence, the $B\to\eta_{c}K$ modes require a more thorough study. We
have predicted the branching ratios, the polarization fractions,
and the relative phases associated with the $B\to
(\chi_{c1},\chi_{c0},\eta_c)K^{*}$ decays, which can be compared
with future measurements.

\vskip 1.0cm We thank K.T. Chao and H.Y. Cheng for useful
discussions. This work was supported by the National Science
Council of R.O.C. under Grant No. NSC-93-2112-M-001-014.





\appendix

\section{FACTORIZATION FORMULAS}

For the $B$ meson wave function, we adopt the model \cite{KLS},
\begin{eqnarray}
\Phi_{B}(x,b)=N_{B}x^{2}(1-x)^{2}\exp \left[ -\frac{1}{2} \left(
\frac{xm_{B}}{\omega _{B}}\right) ^{2} -\frac{\omega
_{B}^{2}b^{2}}{2}\right] \label{bw} \;,
\end{eqnarray}
with the shape parameter $\omega_{B}=0.4$ GeV. The normalization
constant $N_{B}= 91.784$ GeV is related to the decay constant
$f_{B}=190$ MeV (in the convention $f_{\pi}=130$ MeV). The $K$ and
$K^*$ meson distribution amplitudes have been derived from QCD sum
rules \cite{PB1,PB2},
\begin{eqnarray}
\Phi _{K}(x) &=&\frac{3f_K}{\sqrt{2N_{c}}}
x(1-x)\left\{1+0.51(1-2x)+0.3[5(1-2x)^{2}-1]\right\}\;,
\label{kk}\\
\Phi _{K}^{p}(x) &=&\frac{f_K}{2\sqrt{2N_{c}}}\left[
1+0.24C_{2}^{1/2}(1-2x)-0.11C_{4}^{1/2}(1-2x)\right] \;,
\label{kp}\\
\Phi _{K}^{\sigma }(x) &=&\frac{f_K}{2\sqrt{2N_{c}}}(1-2x)\left[
1+0.35(10x^{2}-10x+1)\right] \;, \label{ks} \\
\Phi_{K^* }( x) &=&\frac{3f_{K^*}}{\sqrt{2N_{c}}}x(1-x)
\Big[1+0.57(1-2x)+0.07C^{3/2}_2(1-2x)\Big]\;,
\label{pk2}\\
\Phi_{K^*}^{t}( x) &=&\frac{f_{K^*}^T}{2\sqrt{2N_{c}}} \bigg\{
0.3(1-2x)\left[3(1-2x)^2+10(1-2x)-1\right]+1.68C^{1/2}_4(1-2x)
\nonumber \\
&&+0.06(1-2x)^2\left[5(1-2x)^2-3\right] +0.36\left[
1-2(1-2x)(1+\ln(1-x))\right] \bigg\} \;,
\label{pk3t}\\
\Phi _{K^*}^s( x)  &=&\frac{f_{K^*}^T}{2\sqrt{2N_{c}}} \bigg\{
3(1-2x)\left[1+0.2(1-2x)+0.6(10x^2-10x+1)\right]
\nonumber \\
& &-0.12x(1-x)+0.36[1-6x-2\ln(1-x)]\bigg\} \;,
\label{pk3s}\\
\Phi_{K^*}^{T}(x)&=&\frac{3f_{K^*}^T}{\sqrt{2N_c}} x(1-x)
\Big[1+0.6(1-2x)+0.04C^{3/2}_2(1-2x)\Big]\;,
\label{pkt}\\
\Phi_{K^*}^v(x)&=&\frac{f_{K^*}}{2\sqrt{2N_c}}
\bigg\{\frac{3}{4}\Big[1+(1-2x)^2+0.44(1-2x)^3\Big]
+0.4C^{1/2}_2(1-2x)
\nonumber \\
&& +0.88C^{1/2}_4(1-2x)+0.48[2x+\ln(1-x)] \bigg\}\;,
\label{pkv}\\
\Phi_{K^*}^a(x) &=&\frac{f_{K^*}}{4\sqrt{2N_{c}}}
\Big\{3(1-2x)\Big[1+0.19(1-2x)+0.81(10x^2-10x+1)\Big]\nonumber \\
&&-1.14x(1-x)+0.48[1-6x-2\ln(1-x)]\Big\}\;, \label{pka}
\end{eqnarray}
with the Gegenbauer polynomials
\begin{eqnarray}
C_2^{1/2}(\xi)=\frac{1}{2}(3\xi^2-1)\;,\;\;\;
C_4^{1/2}(\xi)=\frac{1}{8}(35 \xi^4 -30 \xi^2 +3)\;,\;\;\;
C_2^{3/2}(\xi)=\frac{3}{2}(5\xi^2-1)\;,
\end{eqnarray}
and the decay constants $f_K=160$ MeV, $f_{K^*}=200$ MeV, and
$f_{K^*}^{T}=160$ MeV. The coefficients of the Gegenbauer
polynomials correspond to the masses $m_K=0.49$ GeV and
$m_0^K=1.7$ GeV. We adopt the $K^*$ meson mass $m_{K^*}=0.89$ GeV.

\subsection{$B\to (J/\Psi,\chi_{c1}) K^{(*)}$}

We present the spectator amplitudes for the $B\to
(J/\psi,\chi_{c1}) K^{(*)}$ decays below, where the symbol $X$
represents the charmonium $J/\psi$ or $\chi_{c1}$:
\begin{eqnarray}
{\cal M}_{4}^{XK} &=&16\pi C_{F}\sqrt{2N_{c}}%
\int_{0}^{1}[dx]\int_{0}^{\infty }b_{1}db_{1} \Phi
_{B}(x_{1},b_{1})
\nonumber \\
&& \times \Big\{ \Big[ (1-2r^{2}_{2})(1-x_{2}) \Phi _{K} ( x_{3}
)X^{L}(x_{2})
+\zeta_{Y} \frac{1}{2} r^{2}_{2} \Phi_{K}(x_{3})X^{t }(x_{2}) \nonumber \\
&&- r_{K} (1-r^2_2)x_3 \Phi^{p}_{K}(x_3) X^{L}(x_{2})+ r_{K}
\left( 2r^{2}_{2}(1-x_2)+(1-r^2_2)x_3 \right)
\Phi^{\sigma}_{K}(x_3) X^{L}(x_{2})
  \Big]\nonumber \\
&&\times E_{4}(t_d^{(1)})h_d^{(1)}(x_1,x_2,x_3,b_1)
\nonumber \\%
&& - \Big[ (x_2+(1-2r^2_{2})x_{3})\Phi _{K} ( x_{3}
)X^{L}(x_{2})+r^{2}_{2}(2r_{K}\Phi^{\sigma}_{K}(x_{3})-\frac{1}{2}\Phi_{K}(x_{3}))X^{t}(x_{2})
 \nonumber \\
&& -r_{K} (1-r^2_2)x_3 \Phi^{p}_{K}(x_3) X^{L}(x_{2})- r_{K}
\left( 2r^{2}_{2}x_2+(1-r^2_2)x_3 \right) \Phi^{\sigma}_{K}(x_3)
X^{L}(x_{2})\Big]\nonumber \\
&& \times E_{4}(t^{(2)}_d) h_d^{(2)}(x_1,x_2,x_3,b_1) \Big\}\;,
\end{eqnarray}

\begin{eqnarray}
{\cal M}_{6}^{XK} &=&\zeta_{X} 16\pi C_{F}\sqrt{2N_{c}}%
\int_{0}^{1}[dx]\int_{0}^{\infty }b_{1}db_{1} \Phi
_{B}(x_{1},b_{1})
\nonumber \\
&& \times \Big\{  \Big[ (1-x_2+(1-2r^2_{2})x_{3})\Phi _{K} ( x_{3}
)X^{L}(x_{2})+ \zeta_{X}
r^{2}_{2}(2r_{K}\Phi^{\sigma}_{K}(x_{3})-\frac{1}{2}\Phi_{K}(x_{3}))X^{t}(x_{2})
 \nonumber \\
&& -r_{K} (1-r^2_2)x_3 \Phi^{p}_{K}(x_3) X^{L}(x_{2})- r_{K}
\left( 2r^{2}_{2}(1-x_2)+(1-r^2_2)x_3 \right)
\Phi^{\sigma}_{K}(x_3)
X^{L}(x_{2})\Big]\nonumber \\
&&\times E_{6}(t_d^{(1)})h_d^{(1)}(x_1,x_2,x_3,b_1)
\nonumber \\%
&& - \Big[ (1-2r^{2}_{2})x_{2} \Phi _{K} ( x_{3} )X^{L}(x_{2})
+ \frac{1}{2} r^{2}_{2} \Phi_{K}(x_{3})X^{t }(x_{2}) \nonumber \\
&&- r_{K} (1-r^2_2)x_3 \Phi^{p}_{K}(x_3) X^{L}(x_{2})+ r_{K}
\left( 2r^{2}_{2}x_2+(1-r^2_2)x_3 \right) \Phi^{\sigma}_{K}(x_3)
X^{L}(x_{2})
  \Big]\nonumber \\
&& \times E_{6}(t^{(2)}_d) h_d^{(2)}(x_1,x_2,x_3,b_1) \Big\}\;,
\end{eqnarray}

\begin{eqnarray}
{\cal M}_{L4}^{XK^*} &=&16\pi C_{F}\sqrt{2N_{c}}%
\int_{0}^{1}[dx]\int_{0}^{\infty }b_{1}db_{1} \Phi
_{B}(x_{1},b_{1})
\nonumber \\
&& \times \Big\{ \Big[ (1-2r^{2}_{2})(1-x_{2}) \Phi _{K^*} ( x_{3}
)X^{L}(x_{2})
+ \zeta_{X} \frac{1}{2} r^{2}_{2} \Phi_{K^*}(x_{3})X^{t }(x_{2}) \nonumber \\
&&- r_{K^*} (1-r^2_2)x_3 \Phi^{s}_{K^*}(x_3) X^{L}(x_{2})+ r_{K^*}
\left( 2r^{2}_{2}(1-x_2)+(1-2r^2_2)x_3 \right) \Phi^{t}_{K^*}(x_3)
X^{L}(x_{2})
  \Big]\nonumber \\
&&\times E_{4}(t_d^{(1)})h_d^{(1)}(x_1,x_2,x_3,b_1)
\nonumber \\%
&& - \Big[ (x_2+(1-2r^2_{2})x_{3})\Phi _{K^*} ( x_{3}
)X^{L}(x_{2})+r^{2}_{2}(2r_{K^*}\Phi^{t}_{K^*}(x_{3})-\frac{1}{2}\Phi_{K^*}(x_{3}))X^{t}(x_{2})
 \nonumber \\
&& -r_{K^*} (1-r^2_2)x_3 \Phi^{s}_{K^*}(x_3) X^{L}(x_{2})- r_{K^*}
\left( 2r^{2}_{2}x_2+(1-2r^2_2)x_3 \right) \Phi^{t}_{K^*}(x_3)
X^{L}(x_{2})\Big]\nonumber \\
&& \times E_{4}(t^{(2)}_d) h_d^{(2)}(x_1,x_2,x_3,b_1) \Big\}\;,
\end{eqnarray}

\begin{eqnarray}
{\cal M}_{L6}^{XK^*} &=& \zeta_{X} 16\pi C_{F}\sqrt{2N_{c}}%
\int_{0}^{1}[dx]\int_{0}^{\infty }b_{1}db_{1} \Phi
_{B}(x_{1},b_{1})
\nonumber \\
&& \times \Big\{ \Big[ (1-x_2+(1-2r^2_{2})x_{3})\Phi _{K^*} (
x_{3})X^{L}(x_{2})+ \zeta_{X}
r^{2}_{2}(2r_{K^*}\Phi^{t}_{K^*}(x_{3})-\frac{1}{2}\Phi_{K^*}(x_{3}))X^{t}(x_{2})
 \nonumber \\
&& - r_{K^*} (1-r^2_2)x_3 \Phi^{s}_{K^*}(x_3) X^{L}(x_{2})-
r_{K^*} \left( 2r^{2}_{2}(1-x_2)+(1-2r^2_2)x_3 \right)
\Phi^{t}_{K^*}(x_3) X^{L}(x_{2})\Big]
\nonumber \\
&&\times E_{6}(t_d^{(1)})h_d^{(1)}(x_1,x_2,x_3,b_1)
\nonumber \\%
&& - \Big[ (1-2r^{2}_{2})x_{2} \Phi _{K^*} ( x_{3} )X^{L}(x_{2})
+ \frac{1}{2} r^{2}_{2} \Phi_{K^*}(x_{3})X^{t }(x_{2}) \nonumber \\
&&- r_{K^*} (1-r^2_2)x_3 \Phi^{s}_{K^*}(x_3) X^{L}(x_{2})+ r_{K^*}
\left( 2r^{2}_{2}x_2+(1-2r^2_2)x_3 \right) \Phi^{t}_{K^*}(x_3)
X^{L}(x_{2})
  \Big]\nonumber \\
&& \times E_{6}(t^{(2)}_d) h_d^{(2)}(x_1,x_2,x_3,b_1) \Big\}\;,
\end{eqnarray}

\begin{eqnarray}
{\cal M}_{N4}^{XK^*} &=&16\pi C_{F}\sqrt{2N_{c}}
\int_{0}^{1}[dx]\int_{0}^{\infty }b_{1}db_{1} \Phi
_{B}(x_{1},b_{1})
\nonumber \\
&&\times r_{2}\Big\{
\Big[(1-r^2_{2})(1-x_{2})\Psi^{T}_{K^*}(x_3)X^{V}(x_2)
\nonumber \\
&&
 +\zeta_{X} \frac{1}{2}r_{K^*}  ((1+r^{2}_{2}) \Phi^{v}_{K^*}(x_{3})
 -(1-r^{2}_{2})\Psi^{a}_{K^*}(x_{3}) ) X^{T}(x_{2})\Big]
E_{4}(t^{(1)}_d)h_d^{(1)}(x_1,x_2,x_3,b_1)
\nonumber \\
&& + \Big[(1-r^2_{2})x_{2}\Psi^{T}_{K^*}(x_3)X^{V}(x_2)
\nonumber \\
&&
 -\left((1-r^2_{2})\Phi^{T}_{K^*}(x_3)- \frac{1}{2}r_{K^*}
  ((1+r^{2}_{2}) \Phi^{v}_{K^*}(x_{3})+(1-r^{2}_{2})\Psi^{a}_{K^*}(x_{3}) \right)
  X^{T}(x_{2}) \nonumber \\
  && -2r_{K^*}((1+r^{2}_{2})x_{2}+(1-r^{2}_{2})x_{3}) \Phi^{v}_{K^*}(x_{3}) X^{V}(x_2) \Big]
E_{4}(t^{(2)}_d) h_d^{(2)}(x_1,x_2,x_3,b_1) \Big\}\;,
\end{eqnarray}

\begin{eqnarray}
{\cal M}_{N6}^{XK^*} &=&-\zeta_{X} 16\pi C_{F}\sqrt{2N_{c}}
\int_{0}^{1}[dx]\int_{0}^{\infty }b_{1}db_{1} \Phi
_{B}(x_{1},b_{1})
\nonumber \\
&&\times r_{2}\Big\{
\Big[(1-r^2_{2})(1-x_{2})\Psi^{T}_{K^*}(x_3)X^{V}(x_2)
\nonumber \\
&&
 -\zeta_{X} \left((1-r^2_{2})\Phi^{T}_{K^*}(x_3)- \frac{1}{2}r_{K^*}
  ((1+r^{2}_{2}) \Phi^{v}_{K^*}(x_{3})+(1-r^{2}_{2})\Psi^{a}_{K^*}(x_{3}) \right)
  X^{T}(x_{2}) \nonumber \\
  && -2r_{K^*}((1+r^{2}_{2})(1-x_{2})+(1-r^{2}_{2})x_{3})
  \Phi^{v}_{K^*}(x_{3}) X^{V}(x_2) \Big]
  E_{6}(t^{(1)}_d)h_d^{(1)}(x_1,x_2,x_3,b_1)
\nonumber \\
&& + \Big[(1-r^2_{2})x_{2}\Psi^{T}_{K^*}(x_3)X^{V}(x_2)
\nonumber \\
&&
 +\frac{1}{2}r_{K^*}  ((1+r^{2}_{2}) \Phi^{v}_{K^*}(x_{3})-(1-r^{2}_{2})\Psi^{a}_{K^*}(x_{3}) ) X^{T}(x_{2})\Big]
E_{6}(t^{(2)}_d) h_d^{(2)}(x_1,x_2,x_3,b_1) \Big\}\;,
\end{eqnarray}

\begin{eqnarray}
{\cal M}_{T4}^{XK^*} &=& 32\pi C_{F}\sqrt{2N_{c}}
\int_{0}^{1}[dx]\int_{0}^{\infty }b_{1}db_{1} \Phi
_{B}(x_{1},b_{1})
\nonumber \\
&&\times  r_{2}\Big\{ \Big[ (1-x_2)\Phi^{T}_{K^*}(x_3)X^{V}(x_2)
-\zeta_{X} \frac{1}{2} r_{K^*}
\left(\Phi^{v}_{K^*}(x_3)-(1+2r^{2}_{2})\Psi^{a}_{K^*}(x_3)
\right) X^{T}(x_{2}) \Big] \nonumber \\
&&\times E_{4}(t^{(1)}_d)h_d^{(1)}(x_1,x_2,x_3,b_1)
\nonumber \\
&& + \Big[ x_2 \Phi^{T}_{K^*}(x_3)X^{V}(x_2) -
\left(\Phi^{T}_{K^*}(x_3)-\frac{1}{2} r_{K^*}(
\Phi^{v}_{K^*}(x_3)+(1+2r^{2}_{2})\Psi^{a}_{K^*}(x_3) ) \right)
X^{T}(x_{2})  \nonumber
\\
&& -2r_{K^*} ((1+2r^{2}_{2})x_2 +x_{3})\Phi^{a}_{K^*}(x_3)
X^{V}(x_2) \Big]E_{4}(t^{(2)}_d) h_d^{(2)}(x_1,x_2,x_3,b_1)
\Big\}\;,
\end{eqnarray}

\begin{eqnarray}
{\cal M}_{T6}^{XK^*} &=& -\zeta_{X} 32\pi C_{F}\sqrt{2N_{c}}
\int_{0}^{1}[dx]\int_{0}^{\infty }b_{1}db_{1} \Phi
_{B}(x_{1},b_{1})
\nonumber \\
&&\times  r_{2}\Big\{ \Big[ (1-x_2) \Phi^{T}_{K^*}(x_3)X^{V}(x_2)
\nonumber \\
&& -\zeta_{X} \left(\Phi^{T}_{K^*}(x_3)-\frac{1}{2} r_{K^*}(
\Phi^{v}_{K^*}(x_3)+(1+2r^{2}_{2})\Psi^{a}_{K^*}(x_3) ) \right)
X^{T}(x_{2})  \nonumber
\\
&& -2r_{K^*} ((1+2r^{2}_{2})(1-x_2) +x_{3})\Phi^{a}_{K^*}(x_3)
X^{V}(x_2) \Big] E_{6}(t^{(1)}_d)h_d^{(1)}(x_1,x_2,x_3,b_1)
\nonumber \\
&& + \Big[ x_2\Phi^{T}_{K^*}(x_3)X^{V}(x_2) -\frac{1}{2} r_{K^*}
\left(\Phi^{v}_{K^*}(x_3)-(1+2r^{2}_{2})\Psi^{a}_{K^*}(x_3)
\right) X^{T}(x_{2}) \Big] \nonumber\\
&& \times E_{6}(t^{(2)}_d) h_d^{(2)}(x_1,x_2,x_3,b_1) \Big\}\;,
\end{eqnarray}
with $\zeta_{J/\psi}=+1$ and $\zeta_{\chi_{c1}}=-1$ in all the
above factorization formulas.

\subsection{$B\to (\chi_{c0}, \eta_{c}) K^{(*)}$}

We present the spectator amplitudes for the $B\to (\chi_{c0},
\eta_{c}) K^{(*)}$ decays below, where the symbol $X$ represents
the charmonium $\chi_{c0}$ or $\eta_c$:
\begin{eqnarray}
{\cal M}_{4}^{X K} &=&16\pi C_{F}\sqrt{2N_{c}}%
\int_{0}^{1}[dx]\int_{0}^{\infty }b_{1}db_{1} \Phi
_{B}(x_{1},b_{1})
\nonumber \\
&& \times \Big\{ \Big[ (1-x_{2}) \Phi _{K} ( x_{3} )X^v(x_{2})
-\zeta_{X} r^{2}_{2}(2r_{K}\Phi^{p}_{K}(x_3)-\frac{1}{2}  \Phi_{K}(x_{3}))X^{s}(x_{2}) \nonumber \\
&&+ r_{K} (1-r^2_2)x_3 \Phi^{\sigma}_{K}(x_3) X^{v}(x_{2})- r_{K}
\left( 2r^{2}_{2}(1-x_2)+(1-r^2_2)x_3 \right) \Phi^{p}_{K}(x_3)
X^{v}(x_{2})
  \Big]\nonumber \\
&&\times E_{4}(t_d^{(1)})h_d^{(1)}(x_1,x_2,x_3,b_1)
\nonumber \\%
&& - \Big[ (x_2+(1-2r^2_{2})x_{3})\Phi _{K} ( x_{3}
)X^{v}(x_{2})+r^{2}_{2}(2r_{K}\Phi^{p}_{K}(x_{3})-\frac{1}{2}\Phi_{K}(x_{3}))X^{s}(x_{2})
 \nonumber \\
&& -r_{K} (1-r^2_2)x_3 \Phi^{\sigma}_{K}(x_3) X^{v}(x_{2})- r_{K}
\left( 2r^{2}_{2}x_2+(1-r^2_2)x_3 \right) \Phi^{p}_{K}(x_3)
X^{v}(x_{2})\Big]\nonumber \\
&& \times E_{4}(t^{(2)}_d) h_d^{(2)}(x_1,x_2,x_3,b_1) \Big\}\;,
\end{eqnarray}

\begin{eqnarray}
{\cal M}_{6}^{X K} &=& \zeta_{X} 16\pi C_{F}\sqrt{2N_{c}}%
\int_{0}^{1}[dx]\int_{0}^{\infty }b_{1}db_{1} \Phi
_{B}(x_{1},b_{1})
\nonumber \\
&& \times \Big\{  \Big[ (1-x_2+(1-2r^2_{2})x_{3})\Phi _{K} ( x_{3}
)X^{v}(x_{2})- \zeta_{X}
r^{2}_{2}(2r_{K}\Phi^{p}_{K}(x_{3})-\frac{1}{2}\Phi_{K}(x_{3}))X^{s}(x_{2})
 \nonumber \\
&& -r_{K} (1-r^2_2)x_3 \Phi^{\sigma}_{K}(x_3) X^{v}(x_{2})- r_{K}
\left( 2r^{2}_{2}(1-x_2)+(1-r^2_2)x_3 \right) \Phi^{p}_{K}(x_3)
X^{v}(x_{2})\Big]\nonumber \\
&&\times E_{6}(t_d^{(1)})h_d^{(1)}(x_1,x_2,x_3,b_1)
\nonumber \\%
&&  -\Big[ x_{2} \Phi _{K} ( x_{3} )X^{v}(x_{2})
+r^{2}_{2}(2r_{K}\Phi^{p}_{K}(x_3)-\frac{1}{2}  \Phi_{K}(x_{3}))X^{s}(x_{2}) \nonumber \\
&&+ r_{K} (1-r^2_2)x_3 \Phi^{\sigma}_{K}(x_3) X^{v}(x_{2})- r_{K}
\left( 2r^{2}_{2}x_2+(1-r^2_2)x_3 \right) \Phi^{p}_{K}(x_3)
X^{v}(x_{2})
  \Big]\nonumber \\
&& \times E_{6}(t^{(2)}_d) h_d^{(2)}(x_1,x_2,x_3,b_1) \Big\}\;,
\end{eqnarray}

\begin{eqnarray}
{\cal M}_{4}^{X K^*} &=&16\pi C_{F}\sqrt{2N_{c}}%
\int_{0}^{1}[dx]\int_{0}^{\infty }b_{1}db_{1} \Phi
_{B}(x_{1},b_{1})
\nonumber \\
&& \times \Big\{ \Big[ (1-x_{2}) \Phi _{K^*} ( x_{3} )X^{v}(x_{2})
-\zeta_{X} r^{2}_{2}(2r_{K^*}\Phi^{s}_{K^*}(x_3)-\frac{1}{2}  \Phi_{K^*}(x_{3}))X^{s}(x_{2}) \nonumber \\
&&+ r_{K^*} (1-r^2_2)x_3 \Phi^{t}_{K^*}(x_3) X^{v}(x_{2})- r_{K^*}
\left( 2r^{2}_{2}(1-x_2)+(1-r^2_2)x_3 \right) \Phi^{s}_{K^*}(x_3)
X^{v}(x_{2})
  \Big]\nonumber \\
&&\times E_{4}(t_d^{(1)})h_d^{(1)}(x_1,x_2,x_3,b_1)
\nonumber \\%
&& - \Big[ (x_2+(1-2r^2_{2})x_{3})\Phi _{K^*} ( x_{3}
)X^{v}(x_{2})+r^{2}_{2}(2r_{K^*}\Phi^{s}_{K}(x_{3})
-\frac{1}{2}\Phi_{K^*}(x_{3}))X^{s}(x_{2})
 \nonumber \\
&& -r_{K^*} (1-r^2_2)x_3 \Phi^{t}_{K^*}(x_3) X^{v}(x_{2})- r_{K^*}
\left( 2r^{2}_{2}x_2+(1-r^2_2)x_3 \right) \Phi^{s}_{K^*}(x_3)
X^{v}(x_{2})\Big]\nonumber \\
&& \times E_{4}(t^{(2)}_d) h_d^{(2)}(x_1,x_2,x_3,b_1) \Big\}\;,
\end{eqnarray}

\begin{eqnarray}
{\cal M}_{6}^{X K^*} &=&\zeta_{X} 16\pi C_{F}\sqrt{2N_{c}}%
\int_{0}^{1}[dx]\int_{0}^{\infty }b_{1}db_{1} \Phi
_{B}(x_{1},b_{1})
\nonumber \\
&& \times \Big\{  \Big[ (1-x_2+(1-2r^2_{2})x_{3})\Phi _{K^*} (
x_{3}
)X^{v}(x_{2})-\zeta_{X}r^{2}_{2}(2r_{K}\Phi^{s}_{K^*}(x_{3})-\frac{1}{2}\Phi_{K^*}(x_{3}))X^{s}(x_{2})
 \nonumber \\
&& - r_{K^*} (1-r^2_2)x_3 \Phi^{t}_{K^*}(x_3) X^{v}(x_{2})-
r_{K^*} \left( 2r^{2}_{2}(1-x_2)+(1-r^2_2)x_3 \right)
\Phi^{s}_{K^*}(x_3)
X^{v}(x_{2})\Big]\nonumber \\
&&\times E_{6}(t_d^{(1)})h_d^{(1)}(x_1,x_2,x_3,b_1)
\nonumber \\%
&&  -\Big[ x_{2} \Phi _{K^*} ( x_{3} )X^{v}(x_{2})
+ r^{2}_{2}(2r_{K^*}\Phi^{s}_{K^*}(x_3)-\frac{1}{2}  \Phi_{K^*}(x_{3}))X^{s}(x_{2}) \nonumber \\
&&+ r_{K^*} (1-r^2_2)x_3 \Phi^{t}_{K^*}(x_3) X^{v}(x_{2})- r_{K^*}
\left( 2r^{2}_{2}x_2+(1-r^2_2)x_3 \right) \Phi^{s}_{K^*}(x_3)
X^{v}(x_{2})
  \Big]\nonumber \\
&& \times E_{6}(t^{(2)}_d) h_d^{(2)}(x_1,x_2,x_3,b_1) \Big\}\;,
\end{eqnarray}
with $\zeta_{\chi_{c0}}=+1$ and $\zeta_{\eta_{c}}=-1$.

\newpage

\end{document}